\documentclass[
amssymb,
amsmath,
amsfonts,
reprint,
superscriptaddress,
longbibliography,
physrev,
nofootinbib,
aps,
floatfix
]{revtex4-2}

\usepackage{graphicx}
\usepackage{booktabs}
\usepackage[table]{xcolor}
\definecolor{headerblue}{RGB}{62, 92, 125}
\definecolor{lightbluegray}{RGB}{235, 241, 247}
\definecolor{midbluegray}{RGB}{218, 228, 238}
\usepackage[utf8]{inputenc}
\usepackage{graphicx}
\usepackage{comment}
\usepackage{multirow}

\usepackage{dcolumn}
\usepackage[T1]{fontenc}
\usepackage{tabularx}
\usepackage{enumitem}
\usepackage{filecontents}

\usepackage{soul}
\usepackage{caption}
\let\svthefootnote\thefootnote
\newcommand\freefootnote[1]{%
  \let\thefootnote\relax%
  \footnotetext{#1}%
  \let\thefootnote\svthefootnote%
}

\newif\iffastcompile
\fastcompilefalse

\usepackage{graphicx}

\usepackage{xcolor}
\definecolor{color1}{rgb}{0,0.25,0.70}

\usepackage[
  colorlinks=true,
  linkcolor=color1,
  citecolor=color1,
  urlcolor=color1,
  hypertexnames=false
]{hyperref}

\newif\ifshownotes

\shownotestrue

\ifshownotes
    \newcommand{\emb}[1]{\textcolor{cyan}{[EMB: #1]}}
\else
    \newcommand{\emb}[1]{\ignorespaces}
\fi

\newcommand{\hydronium}{$\mathrm{H}_3\mathrm{O}^+$}

\usepackage{tikz}
\usetikzlibrary{arrows.meta,positioning,shapes.geometric,fit,backgrounds}
\definecolor{mygreen}{HTML}{1C5523}
\definecolor{myorange}{HTML}{CF682F}
\definecolor{myblue}{HTML}{587592}
\definecolor{myyellow}{HTML}{DCA940}

\newcommand{\ket}[1]{| #1 \rangle}
\newcommand{\bra}[1]{\langle #1 |}

\begin{document}

\title{Inverse Design of Quantum Control Sequences with Fourier Neural Operators}

\author{Anastasia Pipi}
\affiliation{Department of Physics and Astronomy, University of California, Los Angeles, California 90095, USA}

\author{Valentin Duruisseaux}
\affiliation{Department of Computing and Mathematical Sciences, California Institute of Technology, Pasadena, California 91125, USA}

\author{Emily Been}
\affiliation{Division of Physical Sciences, College of Letters and Science, University of California, Los Angeles, California 90095, USA}

\author{Xuecheng Tao}
\affiliation{Division of Physical Sciences, College of Letters and Science, University of California, Los Angeles, California 90095, USA}

\author{Taylor L. Patti}
\email{tpatti@nvidia.com}
\affiliation{NVIDIA, Santa Clara, California 95051, USA}

\author{Anima Anandkumar}
\affiliation{Department of Computing and Mathematical Sciences, California Institute of Technology, Pasadena, California 91125, USA}

\author{Prineha Narang}
\email{prineha@ucla.edu}
\affiliation{Division of Physical Sciences, College of Letters and Science, University of California, Los Angeles, California 90095, USA}
\affiliation{Electrical and Computer Engineering Department, University of California, Los Angeles, California 90095, USA}

\date{\today}

\freefootnote{

}

\date{\today}
\begin{abstract}

Quantum optimal control is a key tool for steering quantum dynamics, but its computational cost grows rapidly with the Hilbert space dimension. Here, we introduce a Fourier Neural Operator (FNO)-based framework for learning high-dimensional molecular quantum dynamics and accelerating inverse design of control protocols. Given an initial molecular population distribution, laser frequency, and polarization, the FNO predicts the molecular--motional population dynamics up to \(10^7\times\) faster than GPU-accelerated numerical propagation with CUDA-Q
Dynamics. Using this fast and differentiable surrogate, we develop FNO stochastic pulse-measurement planner (FNO-SPMP), a protocol that constructs pulse sequences to purify an initially mixed Boltzmann distribution. We demonstrate the protocol in an 888-dimensional subspace of the hydronium molecule at \(20~\mathrm{K}\), achieving target-state population of \(0.98\) with up to an \(86.2\%\) sequence-success rate. In a shared discrete control space, FNO-SPMP achieves nearly twice the success
rate of a reinforcement-learning baseline while using roughly half as many
quantum-control pulses and reducing pulse-sequence generation time from
approximately $10$ hours to $10{-}20$ minutes. 
These results show that operator-learning surrogates can enable inverse design in quantum systems whose Hilbert spaces are too large for conventional direct optimization.
\end{abstract}

\maketitle
\clearpage

\section{Introduction}

The optimization of control pulses for manipulating quantum systems, known as quantum optimal control, has become an essential tool for building viable quantum technologies
\cite{Koch2022QuantumOptimalControl,SCqubitcontrolShai,SCqubitsMtzoi,trappedionCMonroe,NVsensingrev}. The central goal of control is to tune experimental parameters and steer the dynamics of the system toward a desired outcome. This task is often
tractable for small systems, but becomes increasingly difficult in
high-dimensional Hilbert spaces, where the space of possible control sequences and states grows exponentially.

This challenge is exemplified by polyatomic molecular ions, due to their rich internal structure. Furthermore, the AI-mediated control of these species is well-motivated. This is because, while polyatomic molecular ions' large Hilbert spaces increase their computational complexity, the resulting intricate dynamics also make them promising candidates for various quantum information protocols, e.g., precision measurements at cryogenic temperatures for beyond Standard Model physics research~\cite{safronova_search_2018,demille2024quantum,chiral_mol}. 
Among these ions, hydronium ($\mathrm{H}_3\mathrm{O}^+$) is particularly compelling for such experiments, because its inversion transition frequencies are sensitive to violations of local position invariance, providing a possible probe of dark-energy carriers~\cite{h3o_sensing}.
 
Beyond their role in fundamental physics explorations, molecular ions have also attracted interest as qubits for quantum information processing~\cite{mol_qubits2,mol_qubits}. 
Closely related quantum-information techniques have also transformed molecular-ion spectroscopy. For example, 
quantum logic spectroscopy (QLS) \cite{QLS_1,QLS_2,QLS_3,QLS_4,QLS_5,QLS_6} 
adapts trapped-ion logic operations to 
study such molecular ions by co-trapping the molecule with an atomic logic ion, 
enabling non-destructive readout through their shared motional modes. Before precision spectroscopy can be performed, however, the molecule must be prepared in a well-defined internal quantum state. This requires accurate state tracking and precisely controlled external fields. As the molecular Hilbert space becomes more complex, state preparation in QLS becomes a central challenge.

\begin{figure*}[t]
    \includegraphics[width=0.95\textwidth]{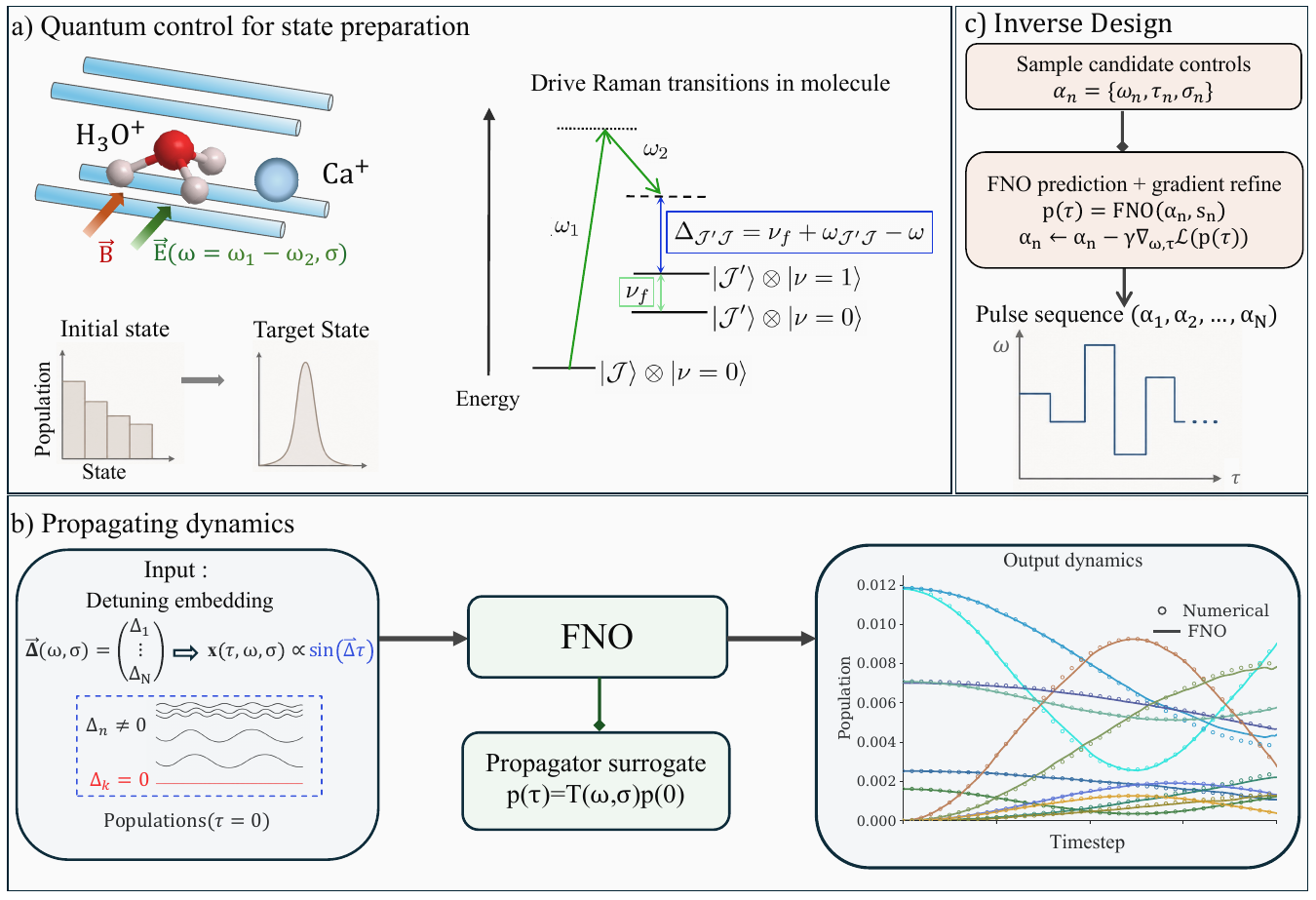}
\caption{\textbf{FNO-based inverse-design pipeline for quantum state preparation.}
\textbf{(a)Quantum control for molecular state preparation.}
Hydronium, \(\mathrm{H_3O^+}\), is co-trapped with a
\(\mathrm{Ca}^{+}\) logic ion in a quantum logic spectroscopy (QLS) setup.
Molecular states \(\mathcal{J}\) and \(\mathcal{J}'\) are addressed by
two-photon Raman sideband pulses driven by fields with angular frequencies
\(\omega_1\) and \(\omega_2\), giving the effective two-photon angular frequency
\(\omega=\omega_1-\omega_2\). The first field is linearly polarized, while the
second has polarization \(\sigma\in\{\sigma^+,\sigma^-\}\), denoted schematically
by \(\vec{E}(\omega,\sigma)\). Appendix~\ref{appx:h3o hami} details how these
electric-field polarizations determine the molecular selection rules. The Raman
sideband drive couples the molecular transition to the shared motional mode,
where \(\nu\) denotes the motional quantum number, through an off-resonant
virtual state indicated by the dotted line. The sideband detuning is \(\Delta_{\mathcal{J}'\mathcal{J}} = \nu_f+\omega_{\mathcal{J}'\mathcal{J}}-\omega\), where \(\nu_f\) is the motional-mode frequency and \(\omega_{\mathcal{J}'\mathcal{J}}\) is the
transition frequency between molecular states \(\mathcal{J}\) and
\(\mathcal{J}'\). Raman pulses are used to prepare a pure molecular state from an
initial thermal mixed state. A magnetic field \(\vec{B}\) lifts the degeneracy of
the hyperfine levels. Energy levels and field directions are schematic and not
to scale.
\textbf{(b)Propagating single-pulse molecular dynamics with the FNO.}
The FNO models the dynamics generated by a single Raman sideband pulse within
the multi-pulse control sequence. At control step \(n\), the model takes as input
the population representation of the current mixed state, shown in the panel as
\(\mathrm{Populations}(\tau=0)\), together with the pulse parameters \(\omega\)
and \(\sigma\). The schematic highlights the detuning-dependent part of the
physics-informed embedding, where sideband transitions produce phase channels
proportional to \(\sin[\boldsymbol{\Delta}(\omega,\sigma)\tau]\). The full input
also includes global drive-frequency information. The FNO learns an effective
population propagator,
\(\mathbf{p}(\tau)=T(\omega,\sigma)\mathbf{p}(0)\), and predicts the
basis-state population trajectories over the local single-pulse propagation
window \(0\leq \tau \leq 4~\mathrm{ms}\) in a single forward pass. The output
panel compares the FNO prediction with numerical simulation.
\textbf{(c) FNO-based inverse design.}
At control step \(n\), the molecule is in the mixed state \(s_n\). A candidate
pulse is described by the general control parameters
\(\alpha=(\omega,\tau,\sigma)\), where \(\omega\) is the effective two-photon
angular frequency, \(\tau\) is the pulse time, and \(\sigma\) is the
polarization of the second electric field. Once a pulse is selected at step
\(n\), these parameters take the chosen values
\(\alpha_n=(\omega_n,\tau_n,\sigma_n)\). For each candidate pulse, the FNO
predicts the population trajectory \(\mathbf{p}(\tau)\) in a single forward
pass, and pulses are selected by minimizing the inverse-design loss. Since the
surrogate is differentiable, the continuous parameters \((\omega,\tau)\) can be
locally refined using gradient-based updates with step size \(\gamma\). Repeating
this procedure produces a control sequence
\((\alpha_1,\ldots,\alpha_N)\) of length \(N\).}
    \label{fig:fig1}
\end{figure*}

Several methods have been developed over the years to design optimized experimental
control parameters. Traditional approaches include gradient-based algorithms
such as GRadient Ascent Pulse Engineering (GRAPE) \cite{GRAPE,GRAPE2,GRAPE3}, and Krotov's method \cite{compareGRAPEKrotov}, as
well as parameterized-pulse methods such as the Chopped RAndom Basis (CRAB) algorithm \cite{CRAB}. Several improved variants of gradient based methods have been developed recently \cite{gradient_based1,gradient_based3,gradient_based4}, but repeated propagation of the quantum state is still required. As a result, these methods can become computationally costly for the large molecular systems considered here.

To address this scalability bottleneck, machine-learning methods have emerged as fast and adaptive tools for quantum optimal control \cite{AI_for_quantum,RL_bukov2018rlcontrol,RL_bukov2018rlcontrol,RL_jelena20,RLpipi, RL_state_prep}. Neural surrogate models have  become increasingly useful for accelerating the propagation of quantum dynamics. In particular, neural quantum states~\cite{NQSreview} have shown that neural networks can compactly represent high-dimensional quantum states, while operator-learning methods~\cite{Zhang24,Zhang25} learn dynamical maps that can reconstruct time evolution. Notably, among neural operators~\cite{kovachki2023neural,azizzadenesheli2024neural,Berner2025Principles}, Fourier neural operators (FNOs)~\cite{li2020fno,valFNOs2026} learn mappings between functions in Fourier space and remain differentiable, making them well suited for predicting time-dependent dynamics in a single forward pass.  This architecture has already been used to accelerate forward propagation in fixed-Hamiltonian spin systems of up to 20 qubits~\cite{FNOsShah}. Subsequent work has extended FNO-based operator learning to several quantum-dynamical settings. These include predicting Floquet Hamiltonians and observable dynamics from experimental data~\cite{FNOfloquet}, as well as learning control-dependent observable dynamics in molecular systems \citep{FNO_pipi} and high-dimensional driven spin systems~\cite{FNO_UNP}.
   
In this work, the molecular-control problem requires expanding the fixed-Hamiltonian propagation framework to incorporate experimentally tunable control parameters. 
Here, we use FNOs in an 888-dimensional molecular subspace not only as surrogate models to propagate observable dynamics, but also for inverse design of molecular state-preparation protocols based on QLS experiments.

The resulting methodology is outlined in Fig. \ref{fig:fig1}. 
We first train the FNO-based surrogate to predict the laser-driven population dynamics of \(\mathrm{H}_3\mathrm{O}^{+}\), achieving high precision throughout the pulse-duration and laser-frequency training domains. By utilizing symmetry relations between polarizations of the laser drives, beginning with a linear $\pi$-pulse and following with either a right $\sigma^+$ or left $\sigma^-$ circularly polarized pulse, the same surrogate can be applied to both polarization combinations without additional training.

Successful training enables the FNO surrogate to accurately evaluate large collections of candidate controls across frequency, polarization, and pulse duration. Finally, we introduce an inverse-design strategy for molecular-state preparation based on an FNO-guided stochastic pulse-measurement planner (FNO-SPMP). Its differentiability further enables optional local refinement of the gradient for selected continuous control parameters.

Here, we make the following contributions:
\begin{enumerate}
    \item To our knowledge, we introduce the first FNO-based
surrogate model for quantum control sequence design. 

    \item We use the FNO surrogate to infer quantum dynamics conditioned on experimental control parameters, predicting full population trajectories for
    candidate Raman sideband pulses.

    \item We introduce the FNO-guided stochastic pulse-measurement planner
    (FNO-SPMP) for quantum control inverse design. The FNO-SPMP scores, selects, and optionally gradient-refines control
    pulses for state preparation.

    \item We demonstrate large-subspace control of hydronium, \hydronium. The FNO surrogate achieves
    up to \(1.84\times 10^7\) speedup relative to direct numerical simulation,
    while FNO-SPMP reduces sequence-generation  approximately
    \(30\text{--}60\times\) compared to state of the art reinforcement learning techniques.
\end{enumerate}

\section{Methods}

We now detail the components of the workflow summarized in
Fig.~\ref{fig:fig1}. We first present the molecular Hamiltonian, the control
parameters, and the state-preparation task. We then describe the training
procedure for the FNO, which is used to predict the basis-state population dynamics
generated by each pulse. Finally, we introduce the FNO-SPMP inverse-design
protocol, which uses the FNO-based surrogate predictions to construct pulse sequences for molecular state preparation.

\label{sub:hami}
\subsection{Hamiltonian}

QLS~\cite{QLS_1} consists of a spectroscopy ion
paired with a logic ion that are confined in the same trap and coupled through their motion via the Coulomb interaction, as shown in Fig.~\ref{fig:fig1}a. The system Hamiltonian is
closely related to the Mølmer--Sørensen entangling gate~\cite{MSgate}. We first
write the QLS Hamiltonian in a general form, where $\ket{\mathcal{J}}$ denotes
an eigenstate of the spectroscopy ion with energy
$E_{\mathcal{J}}=\hbar\omega_{\mathcal{J}}$. The time-independent Hamiltonian is
\begin{equation}
\begin{aligned}
\hat{H}_0
&=
\hat{H}_{\mathrm{spec}} + \hat{H}_{\mathrm{mot}}
\\
&=
\sum_{\mathcal{J}}
\hbar\omega_{\mathcal{J}}
\ket{\mathcal{J}}\bra{\mathcal{J}}
+
\hbar\nu_f
\left(
a^\dagger a + \frac{1}{2}
\right),
\end{aligned}
\end{equation}
where $\nu_f/2\pi=5.164~\mathrm{MHz}$ is the angular frequency of the shared
motional normal mode between the co-trapped ions, and
$a^\dagger$ and $a$ are the corresponding creation and annihilation operators.
The shared motional mode is initialized in its ground state, \(\nu=0\), and can be excited to \(\nu=1\) through a two-photon Raman sideband transition.

In this work, the spectroscopy ion is hydronium, \hydronium, and the logic ion
is \(\mathrm{Ca}^{+}\). We explicitly model the coupled time evolution of the hydronium ion’s internal states and the shared motional mode. The logic ion enters only through
sympathetic cooling and readout. We label each hyperfine-resolved molecular state
of \hydronium by \(\mathcal{J}\). For each rotational manifold \(\ket{J,K}\), where \(J\) is the rotational angular momentum quantum number and \(K\) is the quantum number associated with the projection onto the molecule-fixed symmetry axis, the two opposite-parity
components \(\ket{J,K,p=\pm}\) form an inversion doublet. Here, \(p=\pm\)
denotes the inversion parity. Preparing the molecule in a single internal state
enables precision spectroscopy of opposite-parity inversion transitions, which
are sensitive to variation of the proton-to-electron mass ratio and can
therefore probe fundamental-constant variation relevant to dark-energy
studies~\citep{h3o_sensing}. Further details of the molecular hyperfine
structure are given in Appendix~\ref{appx:h3o hami}.
\begin{figure*}[t]
    \centering
    \includegraphics[width=1\textwidth]{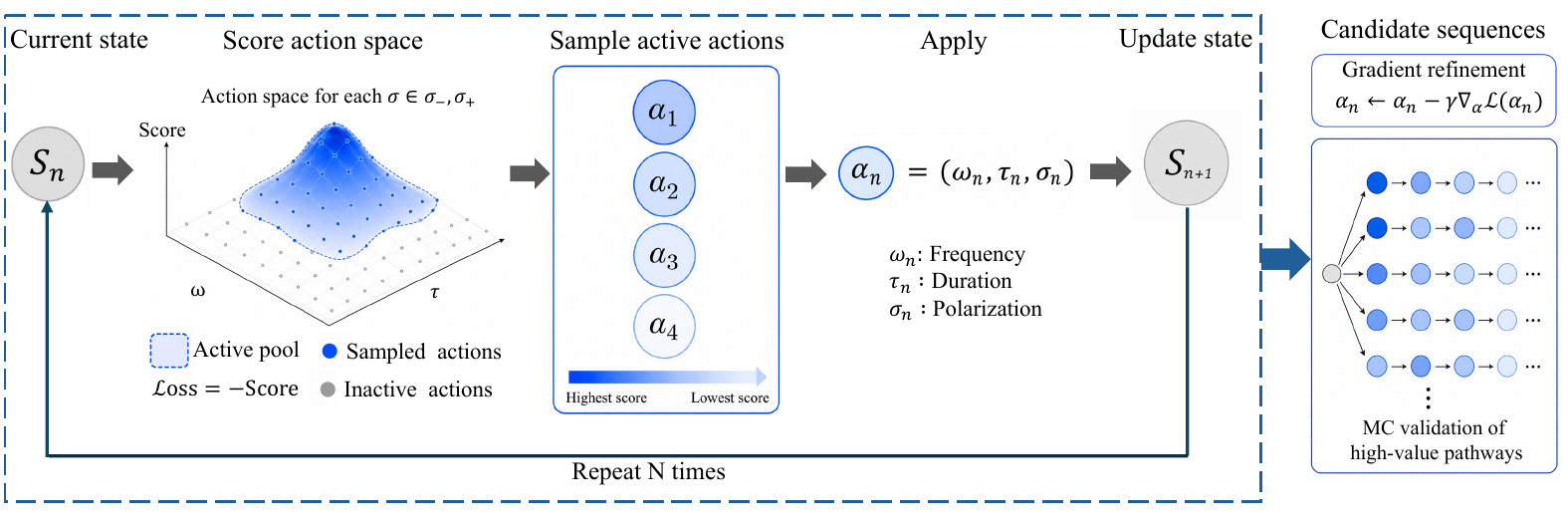}
   \caption{\textbf{Schematic of the stochastic planner (FNO-SPMP) used for inverse design of
  laser-pulse sequences.} Starting from the current state $s_n$, the model scores
  the available actions over the effective laser frequency $\omega$ and time $\tau$, with the
  selection objective taken as the negative score, $\mathcal{L} = -\,\mathrm{Score} = -S(\alpha_n,s_n)$.
  An active pool is then formed from the actions whose score exceeds a defined
  threshold, and for each candidate sequence the planner stochastically samples
  one action from this pool at each pulse step. A sampled action
  $\alpha_n = (\omega_n, \tau_n, \sigma_n)$ (i.e. a pulse specified by its frequency $\omega$,
  duration $\tau$, and polarization $\sigma$) is applied to advance the system
  to the next state $s_{n+1}$. Repeating this for $N$ steps builds up a candidate
  pulse sequence. Each unique candidate can then refined by gradient descent,
  $\alpha_n \leftarrow \alpha_n - \gamma\,\nabla_\alpha \mathcal{L}(\alpha_n)$, on
  its continuous pulse parameters $\alpha_n$ (with step size $\gamma$), and the
  highest-value pathways are validated by Monte Carlo (MC) simulation.}
    \label{fig:inverse_pipeline}
\end{figure*}

The time-dependent Hamiltonian describes a two-photon Raman sideband
pulse, as shown in Fig.~\ref{fig:fig1}a, that couple the spectroscopy
ion's hyperfine states to the shared motional mode:
\begin{equation}
\hat{H}_{\mathrm{int}}
=
\sum_{\mathcal{J}\rightarrow\mathcal{J}'}
\frac{\hbar\Omega_{\mathcal{J},\mathcal{J}'}}{2}
\left[
e^{i[\eta(a+a^\dagger)-\omega t]}
\ket{\mathcal{J}'}\bra{\mathcal{J}}
+
\mathrm{h.c.}
\right],
\end{equation}
where $\Omega_{\mathcal{J},\mathcal{J}'}$ is the effective Rabi coupling
between the hyperfine states $\ket{\mathcal{J}}$ and $\ket{\mathcal{J}'}$. The effective
two-photon angular frequency is $\omega=\omega_1-\omega_2$, set by two electric
fields, $\mathbf{E}_1$ and $\mathbf{E}_2$, with angular frequencies $\omega_1$
and $\omega_2$, respectively. The Lamb--Dicke parameter
$\eta=0.09$ characterizes the coupling strength between molecular and motional degrees of
freedom.

Each Raman pulse is driven by two fields with fixed polarizations. The first
field, \(\mathbf{E}_1\), is \(\pi\)-polarized, while the second field,
\(\mathbf{E}_2\), is chosen to be either \(\sigma^+\)- or
\(\sigma^-\)-polarized. The latter two choices define the two polarization
channels in the control library. We leverage their symmetry to avoid
training separate surrogate models for each channel. For hydronium in its diagonal basis, the \(\sigma^+\) and \(\sigma^-\) channels are related by reversing
the coupling direction between molecular states: for every \(\sigma^+\)
coupling \(\ket{\mathcal{J}}\rightarrow\ket{\mathcal{J}'}\), the corresponding
\(\sigma^-\) coupling drives
\(\ket{\mathcal{J}'}\rightarrow\ket{\mathcal{J}}\) with the same
Rabi-coupling magnitude. The molecular selection rules associated with each electric-field polarization
are detailed in Appendix~\ref{appx:h3o hami}. To resonantly drive a sideband transition between
\(\ket{\mathcal{J}}\) and \(\ket{\mathcal{J}'}\), the two-photon angular
frequency of the Raman drive, \(\omega=\omega_1-\omega_2\), must match the
molecular transition frequency shifted by the shared motional mode frequency.
Thus, the resonant two-photon angular frequencies are
\begin{equation}
\begin{aligned}
\omega^{(+)}
&=
\nu_f + \omega_{\mathcal{J}'\mathcal{J}},
\\
\omega^{(-)}
&=
\nu_f - \omega_{\mathcal{J}'\mathcal{J}}
=
2\nu_f - \omega^{(+)} ,
\end{aligned}
\label{eq:freq_reflection}
\end{equation}
where \(\omega^{(+)}\) and \(\omega^{(-)}\) correspond to the resonant frequencies of the \(\sigma^+\) and
\(\sigma^-\) channels, respectively, \(\nu_f\) is the shared motional mode
angular frequency, and
\(\omega_{\mathcal{J}'\mathcal{J}}
=
\omega_{\mathcal{J}'}-\omega_{\mathcal{J}}\) is the molecular transition
angular frequency.

Therefore, we construct the \(\sigma^-\) surrogate from the trained
\(\sigma^+\) model using the reflection relation in
Eq.~\ref{eq:freq_reflection} and the corresponding state-index permutation.
This avoids training a separate model while preserving the near resonant
couplings. Finally, for an applied Raman pulse with two-photon angular frequency \(\omega\),
we define the detuning from the corresponding sideband resonance as
\begin{equation}
\Delta_{\mathcal{J}'\mathcal{J}}(\omega,\sigma)
=
\omega^{(\pm)}-\omega ,
\label{eq:detuning}
\end{equation}
where \(\sigma\in\{\sigma^+,\sigma^-\}\) denotes the polarization of
\(\mathbf{E}_2\).

\label{sub:Qtask}
\subsection{Quantum Control Task}

In this setting, quantum control for state preparation requires a pulse sequence that steers an initial population distribution toward a single state. We frame this task as an inverse design problem in which we determine the pulse frequencies, durations, and polarizations that maximize the fraction of experimental runs that reach target purity while minimizing protocol duration. 

We define the dynamical process used in the inverse-design stage of the workflow, shown in Fig.~\ref{fig:fig1}c. The objective is to construct a sequence of $N$ control pulses that drive an initially mixed molecular state towards a nearly pure state. We use \(p_{j,n}^{\nu}(\tau,\omega,\sigma)\) to represent the population in the molecular basis state \(\mathcal{J}_j\) and motional manifold \(\nu\) at time
\(\tau\) of a single pulse, with effective two-photon angular frequency \(\omega\) and
polarization \(\sigma\). In the multi-pulse protocol, each control step is indexed by \(n\), and \(\tau\) is reset to zero at the start of each pulse.

We initialize the shared motion in the ground motional manifold, \(\nu=0\), and
set the molecular-state occupation probabilities as described by the Boltzmann distribution
\begin{equation}
p_{j,0}^{0}(\tau=0)
=
\frac{\exp[-E_j/(k_{\mathrm{B}}T_{\mathrm{int}})]}
{\sum_{j'=1}^{M}\exp[-E_{j'}/(k_{\mathrm{B}}T_{\mathrm{int}})]} 
\end{equation}
where \(E_j\) is the energy of molecular basis state \(j\), $M$ denotes the number of molecular states, the temperature
\(T_{\mathrm{int}}\) quantifies the strength of thermal radiation in the molecule at the start of the experiment , and \(k_{\mathrm{B}}\) is the
Boltzmann constant. In the excited motional manifolds, $p_{j,0}^{\nu\neq0}(\tau=0)
=0$. 

The diagonal mixed state 
\begin{equation}
s_n
=
\sum_{j=1}^{M}
p_{j,n}^{0}(\tau=0)
\ket{\mathcal{J}_j,\nu=0}
\bra{\mathcal{J}_j,\nu=0}.
\end{equation} 
denotes the mixed state at the start of control step
\(n\). The purification cycle consists
of the following steps:

\begin{enumerate}
\item \textbf{Applying a control pulse.}
At control step \(n\), the starting mixed state is \(s_n\), and a control pulse
\begin{equation}
\alpha_n
=
(\tau_n,\omega_n,\sigma_n),
\qquad
\sigma_n\in\{\sigma^+,\sigma^-\}
\end{equation}
is applied.
Here, \(\omega_n\) and \(\tau_n\) are the selected values of the continuous
control parameters \(\omega\) and \(\tau\), while the discrete
control \(\sigma_n\) denotes the selected polarization of the second electric field,
\(\mathbf{E}_2\). The pulse drives molecular transitions and partitions the
population into the two motional manifolds.  The basis-state populations evaluated for the selected pulse
\(\alpha_n\) are thus denoted by
\(p_{j,n}^{\nu}(\tau_n,\omega_n,\sigma_n)\), where \(\nu\in\{0,1\}\). In this work, we restrict the motional basis to these two manifolds and omit
higher excited motional states.

\item \textbf{Measuring the motional state.}
At the end of the pulse, \(\tau=\tau_n\), a projective measurement of the
motional state is performed. The probability of observing manifold \(\nu\) is
\begin{equation}
\Pi_{\nu,n}
=
\sum_{j=1}^{M}
p_{j,n}^{\nu}(\tau_n,\omega_n,\sigma_n),
\qquad
\nu\in\{0,1\}.
\label{eq:probs_01}
\end{equation}
The measurement yields a single outcome, which occurs with probability
\(\Pi_{\nu,n}\), for $\nu=0$ or $\nu=1$.

\item \textbf{Motional reset.}
Conditioned on the measured motional outcome, the molecular state is updated by
normalizing the population in the observed motional manifold, followed by
recooling of the shared motion to the ground motional state. The state used to
initialize the next pulse is therefore
\begin{equation}
\begin{gathered}
s_{n+1}
=
\sum_{j=1}^{M}
p_{j,n+1}^{0}(\tau=0)
\ket{\mathcal{J}_j,\nu=0}
\bra{\mathcal{J}_j,\nu=0},
\\[0.8em]
p_{j,n+1}^{0}(\tau=0)
=
\begin{cases}
\displaystyle
\frac{p_{j,n}^{0}(\alpha_n)}{\Pi_{0,n}},
& \text{if } \nu=0, \\[1.0em]
\displaystyle
\frac{p_{j,n}^{1}(\alpha_n)}{\Pi_{1,n}},
& \text{if } \nu=1 ,
\end{cases}
\end{gathered}
\label{eq:meas_out}
\end{equation}
We assume that the measurement and subsequent motional cooling remove coherences,
leading to a population-based description. The protocol continues for either
measurement outcome without post-selection.

\item \textbf{State preparation condition. }
The molecule is considered successfully prepared when one molecular state
dominates the normalized post-measurement population distribution,
\begin{equation}
\max_j
\left[
p_{j,n+1}^{0}(\tau=0)
\right]
\geq
P_{\mathrm{target}}.
\end{equation}
where \(P_{\mathrm{target}}\) is the target preparation threshold. If this
condition is satisfied, the protocol terminates. Otherwise, the conditional
state \(s_{n+1}\) is used for the next control step, and the
pulse--measurement--reset cycle is repeated. %
\end{enumerate}

Selecting an optimal pulse sequence in this setting requires optimization over a
continuous control space. This is computationally expensive because it involves
repeated numerical propagation of the molecular dynamics for many candidate
pulses. Furthermore, since the measurement feedback is the motional state, the
molecular-state dynamics are not directly observed. Tracking the
basis-state populations throughout the pulse sequence is
therefore essential for state preparation in a large state space. To alleviate
these bottlenecks, we train an FNO-based surrogate that replaces each numerical pulse
propagation with a fast, differentiable forward evaluation of the basis-state populations.
\begin{figure*}[t]
    \centering
    \includegraphics[width=1\textwidth]{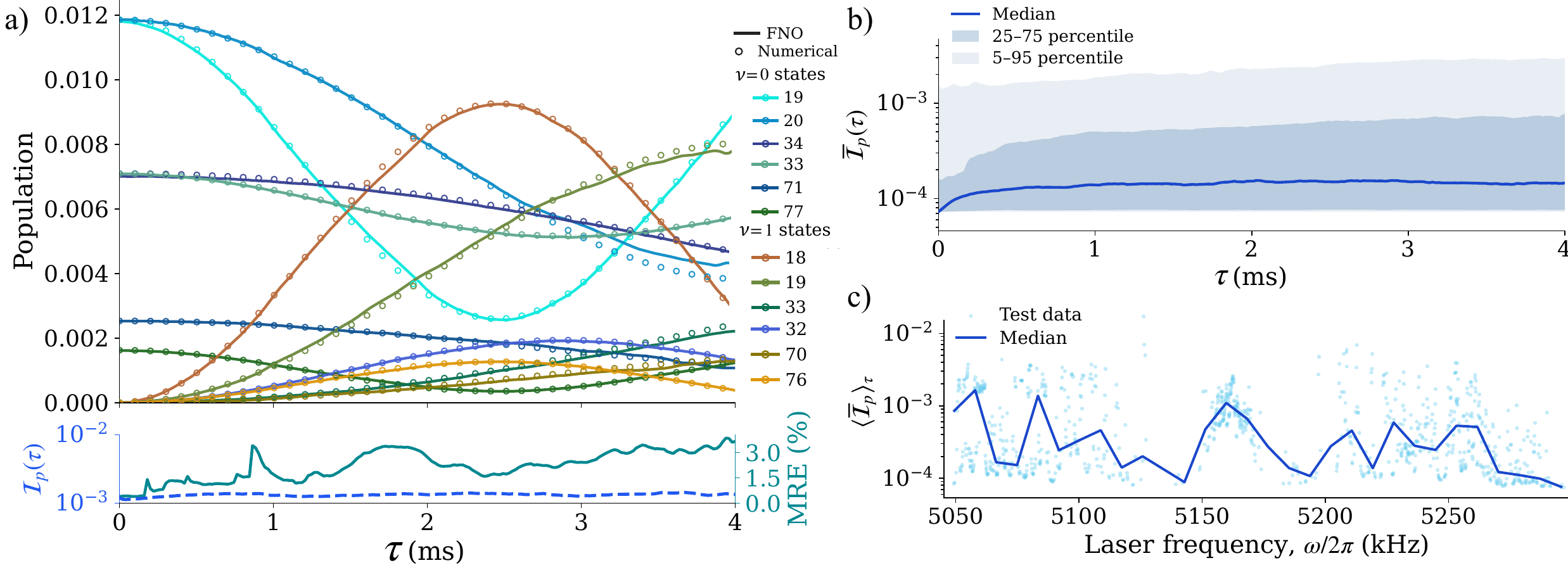}
    \caption{\textbf{Forward-propagation accuracy of the FNO surrogate.}
(a) Basis-state population dynamics predicted by the FNO and compared with CUDA-Q
reference trajectories obtained from the reconstructed unitary propagator. The
trajectory corresponds to a single pulse with drive frequency $\omega/2\pi=5159.6~\mathrm{kHz}$. For clarity, we plot only states with a
population change greater than \(10^{-4}\); approximately stationary states are
omitted from the upper panels but included in the full-state population
infidelity \(\mathcal{I}_p(\tau)\). The lower strip reports
\(\mathcal{I}_p(\tau)=1-\mathcal{F}
_p(\tau)\) over all states and the percentage
mean relative error (MRE) over the plotted active states. States are indexed from lowest to highest energy. 
(b) Time-dependent population infidelity on the test set. For each test
frequency, \(\mathcal{I}_p(\tau)\) is computed for an ensemble of random mixed
initial states and then averaged over the initial-state ensemble, giving one
curve \(\overline{\mathcal{I}}_p(\tau)\) per test frequency. The solid line
shows the median \(\overline{\mathcal{I}}_p(\tau)\) over the test frequencies,
while the shaded regions show the 25--75 and 5--95 percentile ranges across
frequencies at each time point.  (c) Time-averaged population infidelity on the test set,
\(\langle\overline{\mathcal{I}}_p\rangle_{\tau}\), shown as a function of drive
frequency. For each test frequency, the initial-state-averaged curve
\(\overline{\mathcal{I}}_p(\tau)\) is further averaged over the pulse time
\(\tau\), giving one scalar value per frequency. Each point shows this unbinned
value for one test frequency. The solid line shows a binned trend, where
neighboring frequencies are grouped along the frequency axis and the median
\(\langle\overline{\mathcal{I}}_p\rangle_{\tau}\) within each bin is plotted.   }
\label{fig:fno_forward_validation}
\end{figure*}
\subsection{Propagating Dynamics with FNOs}
Machine-learning surrogates provide a route to reducing the computational cost
and limited scalability of traditional numerical integration. In particular,
neural operators (NOs, see Appendix~\ref{appx:NOs}) extend neural networks to
learn mappings between functions~\citep{azizzadenesheli2024neural,Berner2025Principles},
making them well suited for approximating solution operators of dynamical
systems. NOs can approximate broad classes of nonlinear operators~\citep{Kovachki2021_2}
and can act on functions represented on different discretizations, allowing
outputs to be evaluated consistently across resolutions.\\ 

\paragraph{FNO-based surrogate.} We use a Fourier neural operator (FNO)~\citep{li2020fno}, which
parameterizes these operator mappings in Fourier space. The FNO represents the
solution operator through lifting, Fourier, and projection layers
(see Appendix~\ref{appx:FNO-architecture}). The lifting layer maps the input to a higher-dimensional latent representation, the Fourier layers apply learned transformations to a finite set of spectral modes, and the projection layer maps the latent representation back to the desired output space. Because the spectral transformations act globally across the domain, the FNO can efficiently capture long-range structure in time-dependent dynamics. This makes it a computationally efficient and differentiable surrogate for repeated forward prediction and optimization.

The backbone of our pipeline is the FNO-based surrogate, which predicts the time
evolution of the basis-state populations in \hydronium, as shown in
Fig.~\ref{fig:fig1}b. The surrogate model receives as input the mixed state \(s_n\) at the start of control step \(n\), the effective two-photon angular frequency \(\omega\), and the field polarization \(\sigma \in \{\sigma^+,\sigma^-\}\) of \(\mathbf{E}_2\). Here, the surrogate is trained on the \(\sigma^+\) polarization channel. The
dynamics for the \(\sigma^-\) channel are obtained from the same surrogate using
the frequency-reflection relation and state-index permutation described in
Sec.~\ref{sub:hami}.  For each pulse in the multi-pulse sequence, the surrogate predicts the basis-state population trajectories \(\{p_{j,n}^{\nu}(\tau,\omega,\sigma)\}\) over the pulse propagation window \(0 \leq \tau \leq 4~\mathrm{ms}\).

Here, we set \(j=1,\ldots,444\), denoting the lowest- to highest-energy
retained molecular states in \hydronium, respectively, within the \(J=\{1,2,3,4\}\)
rotational manifolds. We include two motional manifolds, \(\nu=0,1\), giving a Hilbert space dimension of \(2\times 444=888\). We construct the surrogate using the block-diagonal Hamiltonian structure
described in Appendix~\ref{appx:block-diagonal}, which decomposes the full
molecular--motional state space into independent dynamical subspaces. A separate
FNO is trained for each block, and the predicted block-level trajectories are
then recombined to recover the full molecular--motional population dynamics. We
now describe the training procedure used for each individual FNO.

We use two main components to enhance model performance:
a physics-informed embedding of input controls and a weighted loss function that prioritizes basis-state populations with significant dynamical changes. \\

\paragraph{Physics-informed embedding.} To build the physics-informed embedding, we encode the applied laser
frequency by its detuning from molecular transitions. The drive couples pairs of molecular states
\(\mathcal{J}\leftrightarrow\mathcal{J}'\). To keep the embedding compact, we include only transitions that are
near resonance with the applied drive and are sufficiently strongly coupled,
\begin{equation}
\left|\Delta_{\mathcal{J}'\mathcal{J}}(\omega,\sigma)\right|
\leq
\Delta_{\max},
\qquad
\left|\Omega_{\mathcal{J}'\mathcal{J}}\right|
\geq
\Omega_{\min},
\end{equation}
where \(\Delta_{\mathcal{J}'\mathcal{J}}(\omega,\sigma)\) is the detuning
defined in Eq.~\ref{eq:detuning},
\(\Omega_{\mathcal{J}'\mathcal{J}}\) is the Rabi coupling for the transition
\(\mathcal{J}\leftrightarrow\mathcal{J}'\), and \(\Delta_{\max}\) and
\(\Omega_{\min}\) are the detuning and coupling cutoffs. We use
\(\Delta_{\max}/2\pi = 10^{4}/(2\pi)~\mathrm{kHz}\) to discard far-detuned terms
that rotate rapidly and average out on the simulation timescale. We use
\(\Omega_{\min}/2\pi = 1/(2\pi)~\mathrm{kHz}\) to remove weak couplings that do
not produce appreciable population transfer within \(4~\mathrm{ms}\).

We index the molecular transitions retained in the embedding by $q=1,...,Q$ with corresponding detunings, $\Delta_q(\omega,\sigma)$. These detunings are
collected into a vector
\begin{equation}
\boldsymbol{\Delta}(\omega,\sigma)
=
\big(
\Delta_1(\omega,\sigma),
\Delta_2(\omega,\sigma),
\ldots,
\Delta_Q(\omega,\sigma)
\big)^{T}.
\end{equation}
Each detuning is embedded as a time-dependent channel,
\begin{equation}
\Phi_q(\tau,\omega,\sigma)
=
s_{\mathrm{emb}}
\frac{
\sin\!\left[\Delta_q(\omega,\sigma)\tau \right]
}{
1+\beta\left|\Delta_q(\omega,\sigma)\right|
},
\end{equation}
where $s_{\mathrm{emb}}=0.05$ rescales the embedding channels to be comparable
in magnitude to the population inputs, while $\beta=0.01$ smoothly suppresses
large-detuning contributions. 

In addition to the detuning channels, we include a global frequency channel. The
applied frequency is normalized over the training frequency range,
\begin{equation}
\omega_{\mathrm{norm}}
=
\frac{\omega-\omega_{\min}}{\omega_{\max}-\omega_{\min}},
\qquad
\tilde{\omega}
=
2\omega_{\mathrm{norm}}-1 .
\end{equation}
where $\omega_{\min}$ and $\omega_{\max}$ are the lower and upper bounds of the FNO training window.  Given that $\tilde{\omega}$ is dimensionless, the global channel is evaluated on
the normalized time coordinate $\tau/\tau_{\max}$,
\begin{equation}
\Phi_{\omega}(\tau,\omega)
=
s_{\mathrm{emb}}
\sin\!\left(2\pi\tilde{\omega}\frac{\tau}{\tau_{\max}}\right),
\end{equation}
where $\tau_{\max}=4~\mathrm{ms}$. This channel provides a smooth global encoding
of the applied frequency across the propagation window. 

For block \(f\), the FNO input is,
\begin{equation}
\mathbf{x}^{(f)}(\tau,\omega,\sigma,\mathbf{s})
=
\left[
\mathbf{s},
\boldsymbol{\Phi}(\tau,\omega,\sigma),
\Phi_{\omega}(\tau,\omega)
\right],
\end{equation}
where
\(\boldsymbol{\Phi}(\tau,\omega,\sigma)
=
(\Phi_1(\tau,\omega,\sigma),\ldots,\Phi_Q(\tau,\omega,\sigma))^T\)
, and
\(\mathbf{s}
=
(p_{1,n}^{0}(\tau=0),\ldots,p_{M_f,n}^{0}(\tau=0))^T\)
is the population-vector representation of the mixed state at the start of control step $n$, and $M_f$ denotes the number of molecular states in the Hamiltonian block. In practice, this channel is evaluated on the
pulse-time grid \(\{\tau^{(r)}\}_{r=1}^{P_\tau}\), with \(P_\tau=200\) time
points over \(\tau\in[0,4]~\mathrm{ms}\). This results to an input size of $[Q+M_f+1,P_\tau]$. This representation allows the model to encode the
oscillatory phase accumulated by each near-resonant transition. The FNO is a trainable operator
\(\mathcal{G}_{\theta}^{(f)}\) that maps this input to the predicted block-level
population dynamics,
\begin{equation}
\mathbf{p}^{(f),\mathrm{pred}}(\tau,\omega,\sigma)
=
\mathcal{G}_{\theta}^{(f)}
\left[
\mathbf{x}^{(f)}(\tau,\omega,\sigma)
\right].
\end{equation}
The predicted molecular--motional population vector for block \(f\) is \(\mathbf{p}^{(f),\mathrm{pred}}
=
(p_{1,n}^{0}(\tau=0),\ldots,p_{M_f,n}^{1}(\tau=0))^T\)
The output therefore has size \([2\times M_f, P_\tau]\), where the factor of two
comes from predicting populations in both motional manifolds, \(\nu=0\) and
\(\nu=1\).

The full predicted molecular--motional population trajectory is obtained by
concatenating the block-level FNO predictions and restoring the original state
ordering,
\begin{equation}
\mathbf{p}^{\mathrm{pred}}(\tau,\omega,\sigma)
=
\mathcal{P}^{-1}
\left[
\bigoplus_{f=1}^{N_{\mathrm{blocks}}}
\mathbf{p}^{(f),\mathrm{pred}}(\tau,\omega,\sigma)
\right].
\label{eq:full_fno_pred}
\end{equation}
Here, \(\bigoplus\) denotes concatenation over the $N_{\mathrm{blocks}}$  Hamiltonian blocks, and
\(\mathcal{P}^{-1}\) maps the block-ordered population vector back to the
original molecular--motional basis. The Hamiltonian block details can be found in
Appendix~\ref{appx:block-diagonal}. To enforce normalization, a softmax
operation is applied over the population channels at each time step.

The embedding combines frequency- and amplitude-modulated features, which have been shown to be beneficial for training FNOs~\citep{valFNOs2026}. Rather than representing the input controls as constant scalar parameters, we lift them into structured oscillatory channels with nonzero spectral content that can be processed directly by the Fourier layers. This provides a richer input representation and improves the expressivity of the surrogate without requiring a deeper network.\\

\paragraph{Activity-weighted loss function.} In addition to the physics-informed embedding, we use an activity-weighted loss
function that gives larger weight to basis-state populations that vary under the
applied control. Each training sample \(b\) corresponds to the basis-state
population trajectories generated from an initial mixed state and a specified
drive setting \((\omega,\sigma)\). We define the activity vector as the largest
population change over the evaluated pulse-time grid,
\begin{equation}
\mathbf{A}_b
=
\max_{r}
\left|
\mathbf{p}^{\mathrm{true}}_b(\tau^{(r)},\omega,\sigma)
-
\mathbf{p}^{\mathrm{true}}_b(\tau^{(1)},\omega,\sigma)
\right| .
\end{equation}
Here, the maximum is taken over the time index \(r=1,\ldots,P_\tau\) and is
applied entrywise to the population vector. Thus, \(\mathbf{A}_b\) is a vector over population components. The reference trajectory $\mathbf{p}_b^{\mathrm{true}}(\tau,\omega,\sigma)$ is computed with
CUDA-Q~\citep{cudaq}. For each sampled control setting
$\alpha=(\tau,\omega,\sigma)$, the full Hamiltonian $H(\tau,\omega,\sigma)$ is used
to evolve the basis states over the full pulse duration $\tau$. This batched
basis-state evolution constructs the corresponding propagator, from which
population trajectories for any diagonal mixed initial state are obtained by
linear recombination. This method is used to generate both the training and test data.

\begin{figure*}[t]
    \centering
    \includegraphics[width=1\textwidth]{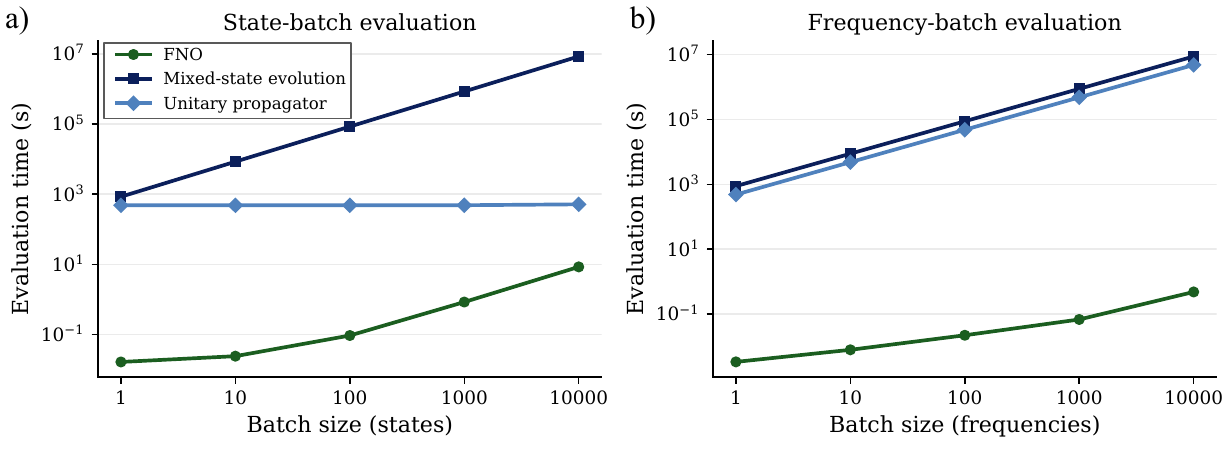}
    \caption{
\textbf{Wall-clock evaluation time for forward propagation using numerical
baselines and the FNO-based surrogate.}
(a) State-batch evaluation, where multiple initial mixed states are propagated
at a fixed drive frequency. (b) Frequency-batch evaluation, where one initial
mixed state is propagated over multiple drive frequencies. We compare three
methods: 1) the FNO surrogate, which predicts the population dynamics in one
batched forward pass, 2) direct mixed-state CUDA-Q evolution, which numerically
propagates the density matrix, and 3) a CUDA-Q unitary-propagator baseline,
where the propagator is reconstructed from batched basis-state evolution and
then applied to the mixed initial state.
}
\label{fig:speedups}
\end{figure*}

\begin{table*}[t]
\centering
\captionsetup{justification=raggedright,singlelinecheck=false}
\caption{
FNO speedup for batched forward evaluation on an NVIDIA A100 80~GB GPU node.
The speedup is defined as
\(T_{\mathrm{CUDA\text{-}Q}}/T_{\mathrm{FNO}}\), so larger values indicate
faster FNO evaluation. Here, ``mixed'' denotes direct mixed-state CUDA-Q
evolution, and ``prop.'' denotes the CUDA-Q unitary-propagator baseline. Bold
entries indicate the maximum speedup within each column.}
\label{tab:speedup_table}
\renewcommand{\arraystretch}{1.15}
\begin{tabular}{ccccc}
\toprule
\multirow{2}{*}{\textbf{Batch size, \(B\)}} &
\multicolumn{2}{c}{\textbf{State-batch speedup}} &
\multicolumn{2}{c}{\textbf{Frequency-batch speedup}} \\
\cmidrule(lr){2-3}\cmidrule(lr){4-5}
&
{\footnotesize Mixed} &
{\footnotesize Prop.} &
{\footnotesize Mixed} &
{\footnotesize Prop.} \\
\midrule
\(10^0\) &
\(5.13\times10^{4}\) &
\(\mathbf{2.92\times10^{4}}\) &
\(2.64\times10^{5}\) &
\(1.45\times10^{5}\) \\
\(10^1\) &
\(3.51\times10^{5}\) &
\(2.00\times10^{4}\) &
\(1.12\times10^{6}\) &
\(6.13\times10^{5}\) \\
\(10^2\) &
\(9.03\times10^{5}\) &
\(5.15\times10^{3}\) &
\(3.98\times10^{6}\) &
\(2.18\times10^{6}\) \\
\(10^3\) &
\(\mathbf{1.00\times10^{6}}\) &
\(5.75\times10^{2}\) &
\(1.30\times10^{7}\) &
\(7.11\times10^{6}\) \\
\(10^4\) &
\(1.00\times10^{6}\) &
\(6.08\times10^{1}\) &
\(\mathbf{1.84\times10^{7}}\) &
\(\mathbf{1.01\times10^{7}}\) \\
\bottomrule
\end{tabular}
\end{table*}
The activity vector is normalized within each trajectory to define the
component-wise weights
\begin{equation}
\mathbf{w}_b
=
1+\lambda
\frac{\mathbf{A}_b}{\max(\mathbf{A}_b)+\epsilon}
\end{equation}
where $\lambda$ sets the strength of the activity weighting, $\epsilon$ is
included for numerical stability, and $\max(\mathbf{A}_b)$ denotes the largest
component of $\mathbf{A}_b$.

For each training sample \(b\), the loss compares the FNO-predicted and true
population trajectories on the pulse-time grid
\(\{\tau^{(r)}\}_{r=1}^{P_\tau}\), with \(P_\tau=200\) points over
\([0,4]~\mathrm{ms}\). At each time point \(\tau^{(r)}\), we compute the
squared error between the predicted and true population vectors,
\(\mathbf{p}^{\mathrm{pred}}_b\) and \(\mathbf{p}^{\mathrm{true}}_b\). This
error is multiplied entrywise by the activity-weight vector \(\mathbf{w}_b\),
averaged over population components, and then averaged over the pulse-time grid:
\begin{equation}
\begin{aligned}
\mathcal{L}
=
\sum_b
\left[
\frac{1}{P_\tau}
\sum_{r=1}^{P_\tau}
\left\langle
\mathbf{w}_b
\left(
\mathbf{p}^{\mathrm{pred}}_b(\tau^{(r)},\omega,\sigma)
\right.\right.\right.
\\
\left.\left.\left.
-
\mathbf{p}^{\mathrm{true}}_b(\tau^{(r)},\omega,\sigma)
\right)^2
\right\rangle_{\mathrm{pop}}
\right]^{1/2}.
\end{aligned}
\end{equation}
Here \(\langle\cdot\rangle_{\mathrm{pop}}\) denotes an average over population components. This weighting penalizes errors more strongly for dynamically active populations while retaining all population components in the loss. Further training details are provided in Appendix~\ref{appx:FNO-hps}.

\subsection{FNO-guided Inverse Design}

We use the accelerated dynamics evaluation enabled by the FNO surrogate to perform efficient quantum control for QLS-based molecular state preparation. The pulse--measurement purification cycle introduced in Sec.~\ref{sub:Qtask} builds on established QLS protocols, where sideband pulses, motional measurements, and state-dependent updates are used for molecular state preparation~\citep{QLS_4, QLS_1, RLpipi}. This purification cycle has also been formulated previously in RL-QLS~\citep{RLpipi}, where a learned policy selects pulses conditioned on the measurement history. However, prior approaches are limited to restricted discrete action spaces and require expensive RL training, while the differentiability of the FNO allows for gradient-based optimization in a continuous space.  

Here, we introduce the FNO-guided stochastic pulse-measurement planner (FNO-SPMP), which reduces pulse-sequence generation cost while enabling exploration of a larger control space with optional gradient-based optimization to refine the continuous control parameters. This makes large-scale pulse sequence optimization computationally tractable. The objective of the FNO-SPMP protocol is to determine a pulse sequence that achieves the state-preparation condition with the minimal number of pulses, utilizing the pulse–measurement–reset cycle described in Sec.~\ref{sub:Qtask}. 

At each control step \(n\), pulse \(\alpha_n\) is selected and applied, chosen from the control landscape $\alpha$. Following the motional measurement at step \(n\), the molecular population
distribution is normalized within the measured motional branch and used to
initialize the next pulse after motional reset. State preparation is considered
successful when this conditional distribution contains a basis-state population
of at least \(P_{\mathrm{target}}\),
\begin{equation}
\max_j
[p_{j,n+1}^{0}(\tau=0)]
\geq
P_{\mathrm{target}}.
\end{equation}
Here, the maximization is taken only over the molecular-state index \(j\), and
\(p_{j,n+1}^{0}(\tau=0)\) denotes the normalized molecular population at the
beginning of the next pulse in the ground motional manifold. Once this condition is met, the protocol terminates. Since the pulse--measurement--reset cycle follows a
stochastic trajectory determined by the measurement outcomes, the planner
in Fig.~\ref{fig:inverse_pipeline} uses branch-aware scores rather than optimizing a single
deterministic trajectory. To build the pulse sequence, we proceed in several steps, as outlined below.\\

\paragraph*{1. Evaluation on a discrete action grid.}

We first define a fixed discrete action grid over
frequency and pulse duration, use the FNO to propagate the dynamics for each candidate control on this grid, so that they can later be scored. Frequencies are sampled within the FNO training window using a
predefined spacing \(\Delta\omega\),
\begin{equation}
\omega_a
=
\omega_{\min}+a\Delta\omega,
\qquad
\omega_a\leq\omega_{\max},
\label{eq:omega_sampling}
\end{equation}
where \(\omega_{\min}\) and \(\omega_{\max}\) are the lower and upper bounds of
the FNO training window, and \(a\) indexes the frequency-grid points. For each
frequency, candidate pulse durations are selected from the fixed FNO time grid,
\begin{equation}
\tau_c
=
\tau_{\min}+c\Delta \tau,
\qquad
\tau_c\leq\tau_{\max},
\label{eq:tau_sampling}
\end{equation}
where \(\tau_{\min}\) and \(\tau_{\max}\) define the range of pulse durations
considered during inverse design, and \(c\) indexes the pulse-duration grid
points.

To construct a pulse sequence, the planner scores candidate controls at each
control step, conditioned on the mixed state \(s_n\) at the start of pulse \(n\),
as outlined in Fig.~\ref{fig:inverse_pipeline}. The FNO performs batched evolution of the
candidate controls by predicting the full population trajectory
\(\mathbf{p}^{\mathrm{pred}}(\tau,\omega_a,\sigma )\), defined in
Eq.~\ref{eq:full_fno_pred}, for each candidate frequency \(\omega_a\) and
polarization \(\sigma\). The candidate duration \(\tau_c\) then selects the
endpoint of this trajectory used to score the control.\\

\paragraph*{2. Scoring and ranking of candidate controls.} Each candidate control, $\alpha$, is scored using three physically motivated criteria:
selective transfer, branch purity, and excited-branch population. We write the
score as
\begin{equation}
\begin{aligned}
S(\alpha,s_n)
={}&
w_{\mathrm{tr}}S_{\mathrm{tr}}(\alpha,s_n)
+
w_{\mathrm{br}}S_{\mathrm{br}}(\alpha,s_n)
\\
&+
w_1(s_n)\Pi_1(\alpha,s_n).
\end{aligned}
\end{equation}
Here \(S_{\mathrm{tr}}\) rewards selective transfer of the most populated
molecular states into the excited motional branch, \(S_{\mathrm{br}}\) rewards
post-measurement population distributions that are concentrated in a single molecular state, and
\(\Pi_1(\alpha,s_n)\) is the probability of observing the excited
motional branch at the end of the candidate pulse. The
\(\Pi_1(\alpha,s_n)\) term encourages measurement-induced
partitioning by favoring candidates that produce a non-negligible \(\nu=1\)
branch probability. The weights \(w_{\mathrm{tr}}\), \(w_{\mathrm{br}}\), and
\(w_1(s_n)\) control the relative importance of these objectives. In the
implementation, \(w_{\mathrm{tr}}=0.5\) and \(w_{\mathrm{br}}=1.5\), while
\(w_1(s_n)\) is chosen adaptively to encourage \(\nu=1\) branch exploration when
the molecular distribution is still far from pure. These weights are fixed
heuristically and are not further optimized. \\ %

\paragraph*{3. Stochastic branch-aware pulse planner.} The candidates are ranked by decreasing score \(S(\alpha,s_n)\). The
active pool contains the top \(N_{\mathrm{pool}}=16\) controls, together with
any additional controls whose scores lie within \(\delta S=0.003\) of the best
score. One pulse is then sampled uniformly from this active pool. Thus,
high-scoring controls are favored through the construction of the pool, while
uniform sampling within the pool keeps the generated sequence stochastic. The
selected pulse, $\alpha_n$, is chosen and applied at step \(n\).

After every pulse, the planner must account for both motional measurement
outcomes, \(\nu=0\) and \(\nu=1\). Constructing the full decision tree is
expensive, because after each pulse the state can be updated conditioned on
either measurement outcome, so the number of possible branches grows
exponentially with the number of pulses. To keep the search tractable, the
planner builds the main pulse sequence by following the \(\nu=0\) outcome at
each control step \(n\). The planning state is therefore updated using the
normalized \(\nu=0\) branch, giving the next state \(s_{n+1}\). This choice is motivated by the structure of the sideband dynamics. The
\(\nu=1\) outcome corresponds to the population transferred into the excited
motional manifold by the selected pulse, which is usually drawn from a smaller
subset of molecular states and is therefore often more concentrated. In
contrast, the \(\nu=0\) outcome contains the remaining untransferred population
and is typically less pure. Following the \(\nu=0\) branch therefore focuses the
main sequence on the harder state-preparation trajectory.

At each step, before advancing along the main \(\nu=0\) branch, the planner also
checks the \(\nu=1\) outcome. If the \(\nu=1\) branch has non-negligible
probability but does not yet satisfy the target threshold
\(P_{\mathrm{target}}\), the planner continues that branch for up to three
additional pulses. These additional pulses account for possible consecutive
\(\nu=1\) measurement outcomes without requiring the full exponentially growing
decision tree. This keeps the search efficient while retaining the measurement
branches most relevant for successful state preparation.

We can generate multiple pulse sequences by repeating the FNO evaluation, scoring, and stochastic sampling of the control parameters many times.\\

\paragraph*{4. (Optional) Gradient-based optimization refinement.} After these sequences are generated, the continuous pulse parameters selected can be
optionally refined further using gradient-based optimization, leveraging the differentiability of the FNO, as shown in the
rightmost box of Fig.~\ref{fig:inverse_pipeline}. For a selected pulse
\(\alpha_n=(\omega_n,\tau_n,\sigma_n)\), the polarization \(\sigma_n\) is kept
fixed, while the continuous parameters \(\omega_n\) and \(\tau_n\) are refined by
gradient descent on the inverse-design loss,
\begin{equation}
(\omega_n,\tau_n)
\leftarrow
(\omega_n,\tau_n)
-
\gamma
\nabla_{(\omega_n,\tau_n)}
\mathcal{L}(\alpha_n,s_n),
\end{equation}
where \(\gamma\) is the gradient step size and
\(\mathcal{L}(\alpha_n,s_n)=-S(\alpha_n,s_n)\). The refined frequency and pulse
duration are constrained to remain within the FNO training window and the allowed
pulse duration range. \\

\paragraph*{5. Monte Carlo validation of designed sequences.} We evaluate each generated pulse sequence using Monte Carlo rollouts on the
truncated decision tree constructed by the planner. Each rollout emulates a
single experimental run. At every pulse, the motional measurement outcome is
sampled according to the population in each branch, \(\Pi_{0,n}\) and \(\Pi_{1,n}\) (Eq. \ref{eq:probs_01}).
The state is then collapsed to the sampled branch and updated using the
corresponding normalized post-measurement population distribution. The rollout
continues by following the pulse sequence stored on that branch of the decision
tree.

Candidate sequences are ranked by the success fraction, defined as the fraction
of rollouts that reach the state-preparation threshold, and by the mean number
of pulses required to reach it. \\

\paragraph*{Control library.}

In this work, the control library includes both \(\sigma^{+}\) and \(\sigma^{-}\) Raman
sideband pulses. Since the FNO is trained on the \(\sigma^{+}\) dynamics, the
corresponding \(\sigma^{-}\) response is obtained using the reflected-frequency
condition in Eq.~\ref{eq:freq_reflection}, together with the state index permutation described in Sec.~\ref{sub:hami}. The surrogate is
evaluated over the experimentally relevant frequency window
\(\omega/2\pi \in [5050,5300]~\mathrm{kHz}\), which contains a dense set of
kHz-separated molecular transitions and can be addressed continuously in
experiment. In addition to these continuously sampled Raman-sideband pulses, the control
library includes two discrete THz transitions that are experimentally
addressable with a frequency-comb drive~\citep{QLS_3}. These THz controls are treated as fixed pulse primitives rather than continuously tunable controls. They are therefore not included in the FNO training domain and are not refined by gradient-based optimization. Instead, because only a small number of such transitions are used, their propagators are computed separately by exact numerical integration and applied directly during candidate scoring. This keeps
the surrogate restricted to the continuous Raman-sideband domain, \(\omega/2\pi \in [5050,5300]~\mathrm{kHz}\), while still allowing the action library to include experimentally available THz transitions that improve connectivity within the molecular subspace.

\section{Results}

\subsection{Accuracy of the FNO Predictions}

We first evaluate the ability of the FNO surrogate to reproduce the
basis-state populations generated by direct numerical propagation with
CUDA-Q. To test generalization, we benchmark the surrogate predictions
$p_j^\nu(\tau,\omega,\sigma)$ on frequencies unseen during training and on random mixed
initial states. The comparison is performed over the training time window,
$0\leq \tau \leq 4~\mathrm{ms}$, sampled at $P_\tau=200$ timesteps,
$\{\tau^{(r)}\}_{r=1}^{P_r}$, which corresponds to the evolution during a
single pulse at one step of the sequence.  
The results are summarized in Fig.~\ref{fig:fno_forward_validation}. The
performance analysis is shown for the \(\sigma^+\) channel, which is the
polarization channel used during training. The \(\sigma^-\)
dynamics are obtained from the trained \(\sigma^+\) surrogate using the
frequency-reflection relation and state-index permutation described in
Sec.~\ref{sub:hami}. 

Because the model predicts
population distributions rather than full quantum states, we quantify accuracy
by population fidelity. The expression for population fidelity at time $\tau$ given a fixed $\omega$ and $\sigma$ is,
\begin{equation}
\mathcal{F}_p(\tau) = \left( \sum_i \sqrt{ p_i^{\mathrm{true}}(\tau,\omega,\sigma)\, p_i^{\mathrm{pred}}(\tau,\omega,\sigma) } \right)^2 
\end{equation}
and we report the corresponding population infidelity,
\begin{equation}
\mathcal{I}_p(\tau) = 1-\mathcal{F}_p(\tau),
\end{equation}
which measures the discrepancy between the FNO predicted and direct numerical populations. 

The FNO-based surrogate accurately captures transitions of varying strength. For a
representative trajectory at \(\omega/2\pi = 5159.6~\mathrm{kHz}\), we plot the
predicted basis-state populations for the dynamically active channels, defined
as those whose population changes by more than \(10^{-4}\), together with the
direct numerical trajectories. The predicted populations closely track the
direct numerical results for the active channels in both motional manifolds, as
shown in Fig.~\ref{fig:fno_forward_validation}a. States with no appreciable
dynamics are omitted on the plot for clarity. The lower panel reports the mean relative
error (MRE) over the plotted states, which remains below \(3.5\%\) throughout
the evolution. For this trajectory, the population infidelity computed over all
basis states and averaged over time,
\(\langle\mathcal{I}_p\rangle_{\tau}
=
\langle 1-\mathcal{F}_p(\tau)\rangle_{\tau}\), lies low at
\(1.03\times 10^{-3}\).

To evaluate the surrogate error over the duration of a single pulse, we compute the population
infidelity \(\mathcal{I}_p(\tau)\) on all test frequencies. For each test
frequency, \(\mathcal{I}_p^{(l)}(\tau)\) is evaluated for
\(L_{\mathrm{init}}=500\) random initial mixed states and then averaged over the
initial-state ensemble,
\begin{equation}
\overline{\mathcal{I}}_p(\tau)
=
\frac{1}{L_{\mathrm{init}}}
\sum_{l=1}^{L_{\mathrm{init}}}
\mathcal{I}_p^{(l)}(\tau).
\end{equation}
This gives one averaged time-dependent infidelity curve,
\(\overline{\mathcal{I}}_p(\tau)\), for each test frequency. We then compare
these averaged curves across the test frequencies. The prediction error remains consistently low throughout the entire pulse-duration window. The median
\(\overline{\mathcal{I}}_p(\tau)\) across test frequencies stays close to
\(10^{-4}\), and the 95th percentile remains below \(2\times10^{-3}\), as shown
in Fig.~\ref{fig:fno_forward_validation}b. The shaded regions show the 25--75
and 5--95 percentile ranges across test frequencies. The consistently low and stable infidelity curves demonstrate that surrogate errors do not
accumulate appreciably over the prediction window.

Comparable accuracy is maintained across the test-frequency range. Using the
same initial state averaged curves from Fig.~\ref{fig:fno_forward_validation}b,
we average over the pulse time \(\tau\) to obtain one scalar error value for
each test frequency,
\begin{equation}
\left\langle \overline{\mathcal{I}}_p \right\rangle_{\tau}
=
\frac{1}{P_\tau}
\sum_{r=1}^{P_\tau}
\overline{\mathcal{I}}_p(\tau^{(r)}).
\end{equation}
Each point in Fig.~\ref{fig:fno_forward_validation}c shows the value of a single test frequency,
while the solid line shows a binned trend obtained by grouping neighboring
frequencies and plotting the median
\(\left\langle \overline{\mathcal{I}}_p \right\rangle_{\tau}\) within each bin.
The time-averaged infidelity is below \(3\times10^{-3}\) for most test
frequencies, with only weak variation in the binned median across frequency. A
few isolated frequencies show larger errors, but these outliers are rare and do
not represent the typical surrogate performance. Collectively, these results demonstrate
that the FNO accurately captures the population dynamics across pulse time,
drive frequency, and initial state.

\subsection{Computational Speedup}

We now quantify the computational advantage of the FNO surrogate for batched forward evaluations. All direct-simulation baselines were run on NVIDIA A100 80~GB GPU and AMD EPYC 7763 CPU nodes using
the CUDA-Q dynamics workflow. Table~\ref{tab:speedup_table} compares the FNO with two CUDA-Q Dynamics baselines: direct mixed-state evolution and a propagator constructed from batched basis-state evolution. The first baseline directly evolves the mixed state density matrix.
Since batching density matrix evolutions did not reduce wall-clock time, the batch cost is estimated as $B$ times the average single mixed-state evolution time. The second baseline constructs a unitary propagator from batched basis-state evolution and obtains mixed-state trajectories by linear recombination. Although the propagator baseline is efficient for state batches, since a single
propagator can be reused at a fixed frequency, the FNO remains substantially
faster. For frequency batches, the numerical propagation cost grows rapidly
because each frequency requires constructing a new propagator.

For a batch of size $B$, the reported speedup is
\begin{equation}
\mathrm{speedup}(B)
=
\frac{T_{\mathrm{CUDA\text{-}Q}}(B)}
{T_{\mathrm{FNO}}(B)} ,
\end{equation}
where $T_{\mathrm{FNO}}(B)$ is the wall-clock time for one batched FNO
evaluation and $T_{\mathrm{CUDA\text{-}}}(B)$ is the corresponding CUDA-Q
baseline time. These wall-clock times are reported in Fig.~\ref{fig:speedups}.

Relative to direct CUDA-Q mixed-state evolution, the FNO speedup increases from $5.13\times10^{4}$ to $1.01\times10^{6}$ for state batches and from $2.64\times10^{5}$ to $1.84\times10^{7}$ for frequency batches as $B$ increases from $10^0$ to $10^4$. Relative to the CUDA-Q propagator baseline, the frequency-batch speedup also grows with $B$, reaching $1.01\times10^{7}$ at $B=10^4$. For state batches, the propagator baseline becomes more competitive at larger $B$ because the same propagator is reused; nevertheless, the FNO remains faster across all tested batch sizes. Together, these results show that the FNO provides a reliable surrogate for
repeated evaluation of candidate pulses,
over a large control space.

\begin{figure*}[t]
    \centering
    \includegraphics[width=\textwidth]{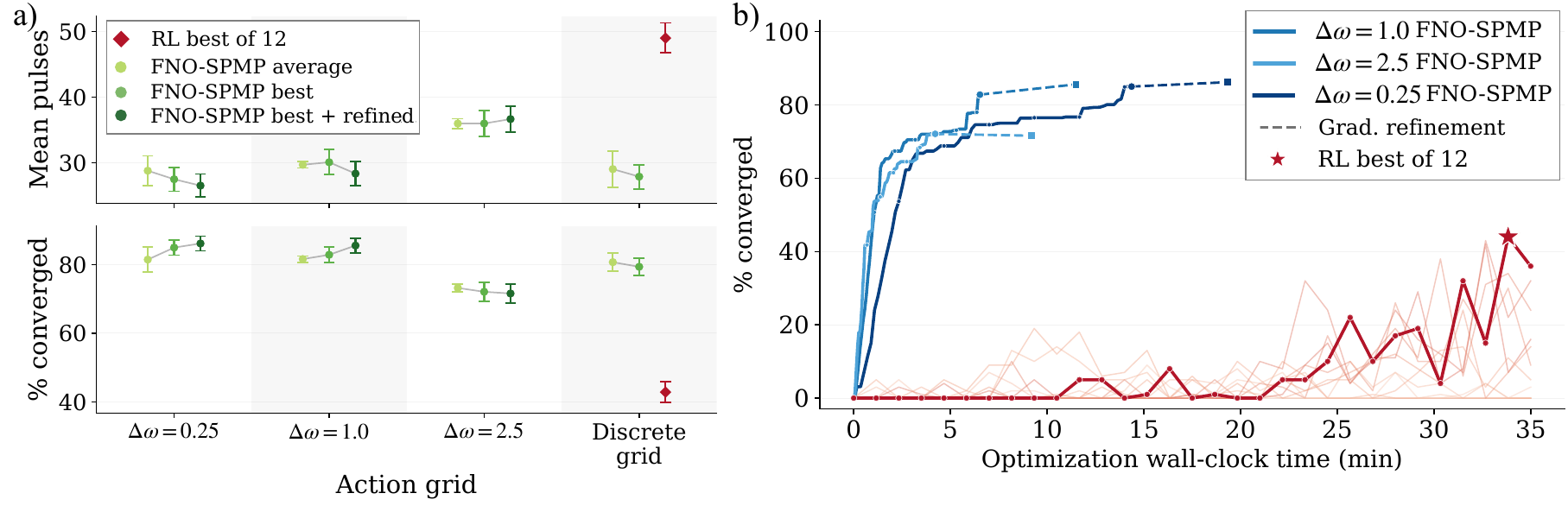}
    \caption{\textbf{Comparison of FNO-SPMP and RL state-preparation protocols.}
    (a) Numerical validation of FNO-SPMP pulse sequences across action grids. The upper panel shows the percentage of Monte Carlo (MC) rollouts that reach the target purity threshold, and the lower panel shows the mean number of pulses over all rollouts. Markers indicate the average performance over generated sequences, the best selected sequence, and the result after gradient-based refinement. For the FNO-SPMP grids, \(\Delta\omega\) is the frequency-grid spacing used during inverse design. The discrete grid is the action space used for the RL comparison, with both frequency and pulse duration restricted to discrete values. 
    (b) Convergence of MC rollouts versus optimization wall-clock time. FNO curves show the performance of sequence prefixes during pulse-sequence construction, with dashed segments denoting gradient-based optimization refinement. RL curves show greedy evaluation of saved policy snapshots across 12 hyperparameter settings. The bold RL curve gives the best-performing setting, and the star marks the selected policy snapshot.}
    \label{fig:inverse_results}
\end{figure*}

\subsection{FNO-based Inverse Design}

\paragraph{Setup.} The inverse-design protocol aims to purify the molecular system from an initial
thermal Boltzmann distribution over its internal states into a single pure state.
Although cryogenic operation improves state-preparation experiments by reducing
the number of thermally occupied hyperfine states, the effective molecular
temperature can still exceed the nominal trap temperature. This can arise from
imperfect thermal shielding and blackbody radiation from experimental
components~\citep{BBR20KChou,BBRtemp}. For example, a recent study reported an internal molecular temperature
approximately twice the nominal trap temperature~\citep{BBR20KChou}. We
therefore use an initial molecular temperature of \(20~\mathrm{K}\) to account
for these imperfections. At this temperature, the 888-dimensional Hilbert space
contains more than \(99\%\) of the thermal population, providing an accurate
representation of the initial distribution without neglecting appreciable
population outside the modeled subspace. To construct each pulse sequence,
we set the purity threshold to \(P_{\mathrm{target}}=0.98\) and allow a maximum
of \(N=80\) pulses. \\

\paragraph{Performance.} The inverse-design performance is summarized in
Fig.~\ref{fig:inverse_results}. All metrics are evaluated by direct numerical
propagation of the selected pulse sequences over 1000 Monte Carlo simulation
runs, ensuring that the validation is independent of accumulated FNO prediction
errors. The left panels compare
performance across different action grids. The first three groups correspond
to FNO-SPMP searches on uniform frequency grids with spacing
$\Delta\omega$, as defined in Eq.~\ref{eq:omega_sampling}. For these grids,
the pulse-duration grid follows Eq.~\ref{eq:tau_sampling}, including all timesteps from
$\tau_{\min}=\tau^{(100)}$, $\tau_{\max}=\tau^{(200)}$. For each grid, we report the
average performance over generated FNO-SPMP sequences, the best performing
sequence, and the result after local gradient refinement using the
differentiable FNO surrogate.

For the finest grid resolution, ($\Delta\omega = 0.25$), the best performing sequence designed by the FNO-SPMP protocol, has convergence improving from 85.0\% to 86.2\% after gradient refinement and the mean number of pulses decreasing from 27.51 to 26.54. A similar trend is observed for ($\Delta\omega = 1.0$), where refinement using gradient-based optimization increases the convergence from 82.9\% to 85.6\% and reduces the mean pulse number from 30.09 to 28.34. Although these gradient refined improvements are modest, they show that local gradient refinement can further polish sequences initialized by the discrete FNO-SPMP.

In contrast, the coarser grid ($\Delta\omega = 2.5$) leads to a noticeable reduction in performance. This suggests that, the local refinement alone cannot fully compensate for the poorer initialization. Across all grids, the average performance over the 10 generated candidate sequences remains close to that of the selected best sequence, indicating that the FNO-SPMP reliably produces high-quality protocols even from a relatively small candidate pool.

We have therefore shown that the protocol prepares the target outcome with probability >0.98 in more than 80\% of attempts, while explicitly tracking population across a large molecular subspace.  The remaining unconverged cases are expected in this high-dimensional setting, where residual population can become diffuse across many states and additional pulses provide only marginal gains while increasing exposure to decoherence. These results are a step toward experimentally viable state-preparation protocols that retain and use population pathways through higher-lying states, which can contribute to the final target-state purity.\\

\paragraph{Comparison with RL-based design.} To quantify the improvement gained by using FNOs for this quantum optimal
control task, we compare FNO-SPMP with the reinforcement-learning (RL) baseline
, RL-QLS, from Ref.~\citep{RLpipi} using
the same discrete control space (labeled as the discrete grid in
Fig.~\ref{fig:inverse_results}a). The allowed pulse durations are
\(\{\tau^{(150)},\tau^{(155)},\ldots,\tau^{(195)}\}\), and the frequencies are
randomly sampled over the same frequency range used by FNO-SPMP rather than
specified by a uniform spacing \(\Delta\omega\). Here, the comparison is made between the best RL sequence and the best performing
FNO-SPMP sequence, both constructed from the same action space.
Within this shared grid, FNO-SPMP reaches \(79.4\%\) convergence with a mean of
\(27.9\) pulses, compared with RL's \(42.8\%\) convergence and  larger mean pulse
count of 49.0. This gap indicates that, in this large discrete control space,
RL struggles to learn an effective policy, while the FNO-guided planner can
more efficiently identify useful pulse sequences. 

Beyond final convergence, the two approaches differ in how efficiently they navigate this large time--frequency control space, as shown in Fig.~\ref{fig:inverse_results}b. For the FNO-SPMP, each point corresponds to numerical validation of the sequence prefix obtained after each pulse-selection step, tracing how performance improves as the protocol is constructed. The dashed segments mark the gradient-based optimization refinement step applied once the discrete sequence has been selected. Comparing grid resolutions, $\Delta\omega = 1.0$ reaches a comparable final convergence as the finer $\Delta\omega = 0.25$ grid but does so noticeably faster, as the coarser action space is cheaper to search. The coarser $\Delta\omega = 2.5$ grid is even faster, but converges to a lower value. The intermediate resolution thus offers the best accuracy--cost trade-off.

For the RL baseline, each model is trained for $3000$ episodes as part of a
12-point hyperparameter sweep. Policy snapshots are saved throughout training
and evaluated greedily, and the strongest policy is selected across both
training episode and hyperparameter setting. Further details of the RL training
procedure are provided in Appendix~\ref{appx:RL}. Even with this selection procedure, the best RL policy remains well below
the FNO-SPMP results. The FNO planner reaches high convergence after only a few
sequence-construction steps, whereas the RL policy improves slowly and does not close
the performance gap over the training window. The early success of the FNO-SPMP can be attributed to the inverse-design loss function. The strongest RL policy is
obtained only after the full sweep, requiring approximately $7.5$ hours of
training across the 12 hyperparameter settings and an additional
$\sim 2.5$ hours to validate saved policy snapshots. By contrast, the FNO
result is selected from only ten independent planner runs. As shown in
Fig.~\ref{fig:inverse_results}a, the average FNO performance lies only slightly
below the best selected sequence, indicating that even a single FNO-SPMP run
typically yields a high-performing protocol. Thus, the FNO surrogate improves
both the quality of the designed pulse sequences and the practical efficiency
with which they are found. Finally, FNO-SPMP reduces pulse-sequence generation time by \(30\)--\(60\times\). Depending on the grid spacing \(\Delta\omega\), FNO-SPMP generates a high-performing protocol in approximately \(10\)--\(20\) minutes, compared with about \(10\) hours for the reinforcement-learning baseline to identify a moderately performing protocol.

\section{Conclusion}

In summary, we have introduced a workflow that employs an FNO-based surrogate model to propagate the population dynamics of \hydronium within a large, hyperfine-resolved Hilbert space. The surrogate closely reproduces population dynamics generated by numerical propagation with CUDA-Q Dynamics while achieving speedups of up to
\(10^{7}\times\). Its accuracy is validated across pulse time, drive frequency,
and random mixed initial states, with average population infidelities consistently
below \(3\times10^{-3}\). This advancement renders large-scale control sequence searches computationally feasible.

Building on this accelerated forward model, we develop FNO-SPMP, an FNO-guided stochastic pulse measurement planner for molecular state preparation. The planner constructs pulse sequences that purify a thermal molecular distribution into a single molecular state. All selected sequences are validated through direct numerical propagation over Monte Carlo simulation runs. Across the tested action grids, FNO-SPMP identifies high-performing protocols and further refines selected controls using gradient-based optimization, leveraging the differentiability of the surrogate. It substantially outperforms the RL baseline within a shared discrete control space. Notably, this improvement extends beyond final convergence probability and mean pulse number. FNO-SPMP constructs effective sequences within minutes, whereas RL requires hours of training and policy validation to achieve its best-performing policy.

FNO-based surrogates can therefore be employed not only to accelerate the forward propagation of quantum dynamics and track population evolution in large Hilbert spaces, but
also to accelerate quantum control tasks. At the scale of
molecular QLS systems, direct numerical searches over continuously tunable
control parameters are computationally prohibitive. This bottleneck is central
to many quantum control tasks in large Hilbert spaces, where each candidate
pulse requires costly state evolution. To our knowledge, we introduce the first
FNO-based surrogate framework for quantum control sequence design, enabling continuous optimization that was previously
computationally infeasible.

Through this work, we take a step toward combining operator-learning
frameworks with optimal quantum control in high-dimensional systems. Beyond
the molecular QLS setting studied here, this approach could be extended to
data-driven pulse design in quantum computing and quantum sensing, where noisy
hardware, calibration drift, leakage, and decoherence make both robust control
and exact simulation challenging. More broadly, the ability of FNOs to learn a
wide range of quantum dynamics, suggests a promising route toward controlling
systems that are too large for conventional direct optimization.

\section*{acknowledgments}

The authors acknowledge David R. Leibrandt, Asad Contractor, Arianna Wu, Nick Lackmann, and Byoungwoo Kang for helpful discussions of various aspects of this work. This work is supported by the Department of Energy (DOE) Office of Science (SC) under DE-FOA-0003432 and by Grant No GBMF12976 of the Gordon and Betty Moore Foundation.

\bibliographystyle{unsrtnat}
\bibliography{bib}

\appendix

\section{Hamiltonian} 

\subsection{Hyperfine Hamiltonian} \label{appx:h3o hami}

In this work, the spectroscopy ion is hydronium, \hydronium, and the logic ion
is \(\mathrm{Ca}^{+}\). We only model the dynamics of the hydronium internal states and the shared
motional mode. The logic ion enters only through sympathetic cooling and
readout.  We label each molecular hyperfine state of \hydronium by
\begin{equation}
\mathcal{J} \equiv \{J,K,p,m_F,\xi\},
\end{equation}
where $J$ is the total rotational angular momentum quantum number, $K$ is its projection onto the molecule-fixed symmetry axis, $p=\pm$ is the inversion
parity, and $m_F$ is the projection of the total angular momentum onto the laboratory-frame quantization axis. The index $\xi$ distinguishes hyperfine eigenstates within the same
$\{J,K,p,m_F\}$ manifold, ordered by increasing energy. For each rotational
state $\ket{J,K}$, the two parity components $\ket{J,K,p=\pm}$ form an
inversion doublet. Preparing the molecule in a single internal state enables
precision spectroscopy of opposite-parity inversion transitions, which are
sensitive to variation of the proton-to-electron mass ratio and can therefore
provide probes of fundamental-constant variation relevant to dark-energy
studies~\citep{h3o_sensing}. Further details of the molecular hyperfine
structure are given in Appendix~\ref{appx:h3o hami}.

The molecular Hamiltonian of $\mathrm{H}_3\mathrm{O}^+$ is defined as :
\begin{equation}
\hat{H}_{\mathrm{mol}}
=
\hat{H}_{\mathrm{inv\text{-}rot}}
+
\hat{H}_{\mathrm{Zeeman}}
+
\hat{H}_{\mathrm{s\text{-}r}},
\end{equation}
where $\hat{H}_{\mathrm{inv\text{-}rot}}$ describes the inversion--rotational
structure, $\hat{H}_{\mathrm{Zeeman}}$ accounts for nuclear and rotational
Zeeman shifts, and $\hat{H}_{\mathrm{s\text{-}r}}$ describes nuclear
spin--rotation coupling. A magnetic field $\mathbf{B}$ of $0.36~\mathrm{mT}$ is applied to
lift the degeneracy of the $m_F$ sublevels. At this field strength, the Zeeman
interaction is comparable in energy scale to the spin--rotation coupling. The computational details of
the energy levels and coupling rates of $\mathrm{H}_3\mathrm{O}^+$ are
reported in a subsequent article \cite{H3OWu}.

The molecular eigenstates are labeled by
\begin{equation}
\mathcal{J} \equiv \{J,K,p,m_F,\xi\},
\end{equation}
where $J$ and $K$ are rotational quantum numbers, $p$ is the inversion parity,
$m_F$ is the projection quantum number of the total angular momentum $F$, and
$\xi$ indexes states with the same values of $\{J,K,p,m_F\}$ in order of
ascending energy. The energy of state $\ket{\mathcal{J}}$ is denoted
$\hbar\omega_{\mathcal{J}}$.

The allowed Raman couplings are restricted by the electric-dipole-induced
two-photon selection rules
\begin{equation}
\Delta J = 0,\pm 2,
\qquad
\Delta K = 0,
\qquad
p_{\mathcal{J}} = p_{\mathcal{J}'} .
\end{equation}

In this molecule we can drive transitions using laser fields that couple the different molecular states in the subspace. For each allowed transition $\ket{\mathcal{J}}\rightarrow\ket{\mathcal{J}'}$,
the effective two-photon Raman Rabi rate is computed by summing over the
intermediate excited states $\ket{\mathcal{M}}$,
\begin{equation}
\begin{aligned}
\Omega_{\mathcal{J},\mathcal{J}'}
&=
\frac{1}{4\hbar^2}
\sum_{\mathcal{M}}
\left[
\frac{A_{\mathcal{J}\mathcal{M}\mathcal{J}'}}{
\omega_{\mathcal{J}\mathcal{M}}-\omega_1}
+
\frac{B_{\mathcal{J}\mathcal{M}\mathcal{J}'}}{
\omega_{\mathcal{J}\mathcal{M}}+\omega_2}
\right],
\end{aligned}
\end{equation}
\begin{equation}
\begin{aligned}
A_{\mathcal{J}\mathcal{M}\mathcal{J}'}
&=
\bra{\mathcal{J}'}\mathbf{d}\cdot\mathbf{E}_2\ket{\mathcal{M}}
\bra{\mathcal{M}}\mathbf{d}\cdot\mathbf{E}_1\ket{\mathcal{J}},
\\
B_{\mathcal{J}\mathcal{M}\mathcal{J}'}
&=
\bra{\mathcal{J}'}\mathbf{d}\cdot\mathbf{E}_1\ket{\mathcal{M}}
\bra{\mathcal{M}}\mathbf{d}\cdot\mathbf{E}_2\ket{\mathcal{J}} .
\end{aligned}
\end{equation}
where $\mathbf{E}_1$ and $\mathbf{E}_2$ are the two laser fields with
frequencies $\omega_1$ and $\omega_2$, $\mathbf{d}$ is the molecular dipole
operator, and
\begin{equation}
\omega_{\mathcal{J}\mathcal{M}}
=
\frac{E_{\mathcal{M}}-E_{\mathcal{J}}}{\hbar}
\end{equation}
is the transition frequency between the initial molecular state
$\ket{\mathcal{J}}$ and the intermediate state $\ket{\mathcal{M}}$. The two
terms account for the two possible time orderings of the Raman process.

The field amplitudes are normalized such that the reference transition
\begin{equation}
\begin{aligned}
\{J,K,p,m_F,\xi\}_{i}
&=
\left\{2,2,-,\frac{3}{2},1\right\},
\\
\{J,K,p,m_F,\xi\}_{f}
&=
\left\{2,2,-,\frac{5}{2},1\right\},
\end{aligned}
\end{equation}
has Rabi rate
\begin{equation}
\Omega_{\mathrm{ref}}/ 2\pi
=
2.000 ~\mathrm{kHz}.
\end{equation}
The two fields driving each pulse have fixed polarizations: \(\mathbf{E}_1\) is
\(\pi\)-polarized, while \(\mathbf{E}_2\) is chosen to be either
\(\sigma^+\)- or \(\sigma^-\)-polarized. The allowed transitions satisfy
\begin{equation}
\Delta m_F
=
m_{F,\mathcal{J}'} - m_{F,\mathcal{J}}
=
\begin{cases}
+1, & \sigma^+,\\
-1, & \sigma^- .
\end{cases}
\end{equation}
The two polarization channels are therefore related by reversing the sign of the
allowed \(\Delta m_F\) transition. Changing from \(\sigma^+\) to \(\sigma^-\)
reverses the angular-momentum coupling between molecular states, i.e., for every
\(\sigma^+\) coupling
\(\ket{\mathcal{J}}\rightarrow\ket{\mathcal{J}'}\), the corresponding
\(\sigma^-\) coupling connects
\(\ket{\mathcal{J}'}\rightarrow\ket{\mathcal{J}}\) with the same
Rabi-coupling magnitude.
\subsection{Block Diagonal form} \label{appx:block-diagonal}
The control subspace includes a drive-frequency in the \(\omega/2\pi \in [5050,5300]~\mathrm{kHz}\) range. Within this frequency range, we can neglect couplings that arise from THz-scale transitions, using the rotating-wave approximation. Within the defined MHz frequency range and the selection rules of two-photon transitions, the Hamiltonian becomes block-diagonal :
\begin{equation}
\hat{H}
=
\bigoplus_{b=1}^{N_{\mathrm{blocks}}}
\hat{H}^{(b)},
\end{equation}
where each block $\hat{H}^{(b)}$ contains molecular states connected by the
allowed Raman couplings. States in different blocks are not directly coupled by the controls used in this work.

In this setting we use a truncated basis set of 444 hyperfine-resolved internal states of $\mathrm{H}_3\mathrm{O}^+$, corresponding to the states in $J = \{1,2,3,4\}$. We also account for 2 motional states, the ground and first excited motional states, resulting in a total subspace of dimension $2\times 444 = 888$. 

In this subspace, the  Hamiltonian decomposes into 10 blocks with dimensions
\begin{equation}
\{128,128,120,96,96,84,84,80,36,36\}.
\end{equation}

Some blocks are symmetry-related and have equivalent dynamics up to relabeling of the molecular states. The following block indices are related through state relabeling : $(0,1)$, $(3,4)$, $(5,6)$, and
$(8,9)$.
We therefore train one FNO surrogate for each unique block,
\begin{equation}
b \in \{0,2,3,5,7,8\},
\end{equation}
and use them to reconstruct the full dynamics.

We verified the block decomposition by comparing direct propagation under the
full Hamiltonian with independent propagation of the individual blocks followed
by recombination. The two procedures agree to numerical precision, confirming
that the block-diagonal representation preserves the dynamics. This justifies training and evaluating the FNO
surrogates block-by-block rather than over the full molecular Hilbert space.

\section{Fourier Neural Operator (FNO) for Forward Modeling}

\subsection{Neural Operators}
\label{appx:NOs}

Machine learning models have been developed to overcome the high computational costs and limited scalability of traditional numerical integrators across scientific fields. Among them, standard neural networks map between finite-dimensional vectors, but PDEs define relationships between infinite-dimensional functions. Neural operators extend the neural network paradigm to learn mappings between these functions~\citep{azizzadenesheli2024neural,Berner2025Principles}, making them particularly well-suited for approximating PDE solution operators. They have a universal approximation property for nonlinear operators~\citep{Kovachki2021_2}, accept input functions on any discretization, and output functions that can be queried consistently at arbitrary resolutions. This flexibility improves data efficiency by supporting training with data of varying discretization and fidelity. Various neural operators like the Fourier Neural Operator (FNO)~\citep{li2020fno} have been applied successfully to diverse problems~\citep{gopakumar2023fourier,Kurth2022,Wen2023,Ghafourpour2025NOBLE}. Additionally, physical laws can be incorporated as auxiliary losses terms to enhance or replace supervised data, as demonstrated in physics-informed neural operator frameworks~\citep{li2024pino,lin2025mGNO,ganeshram2025fcpinohighprecisionphysicsinformed}. With these capabilities, neural operators enable efficient inverse design, both through extensive parameter space sampling or gradient-based optimization (enabled by the differentiability of neural operators).

\begin{figure}[t]
\centering
\begin{tikzpicture}[
    >={Stealth[length=3mm]},
    node distance=8mm,
    circnode/.style={circle, draw=black, minimum size=11mm, font=\itshape},
    blue/.style={circnode, fill=myblue!70!blue!30},
    orange/.style={circnode, fill=myorange!80!orange!40},
    red/.style={circnode, fill=myyellow!50},
    yellowcirc/.style={circnode, fill=mygreen!80!green!50}, %
    box/.style={rectangle, draw=black, fill=mygreen!15, rounded corners,
                minimum width=32mm, minimum height=10mm},
    myarrow/.style={->, thick},
    wave/.style={thin, black!70},
]

\node[orange]   (a)    {a(x)};
\node[blue] (P)    [below=of a]  {P};
\node[box]    (f1)   [below=of P]  {Fourier layer 1};
\node[box]    (f2)   [below=of f1] {Fourier layer 2};
\node[]       (dots) [below=of f2] {$\vdots$};
\node[box]    (fT)   [below=of dots] {Fourier layer T};
\node[blue] (Q)    [below=of fT] {Q};
\node[orange]   (u)    [below=of Q]  {u(x)};

\draw[myarrow] (a)  -- (P);
\draw[myarrow] (P)  -- (f1);
\draw[myarrow] (f1) -- (f2);
\draw[myarrow] (f2) -- (dots);
\draw[myarrow] (dots) -- (fT);
\draw[myarrow] (fT) -- (Q);
\draw[myarrow] (Q)  -- (u);

\begin{scope}[xshift=4cm, yshift=-1cm]

    \node[orange] (v) {v(x)};

    \node[yellowcirc] (F)    [below=of v]  {$\mathcal{F}$};
    \node[yellowcirc] (R)    [below=20mm of F]  {R};
    \node[yellowcirc] (Finv) [below=12mm of R]  {$\mathcal{F}^{-1}$};

    \begin{scope}[shift={([xshift=-0mm,yshift=-8mm]F.center)}]
        \draw[wave] plot[domain=-0.75:0.75,samples=50]
            (\x, {0.15*sin(\x*360*0.5/0.75)});
        \draw[wave] plot[domain=-0.75:0.75,samples=50]
            (\x, {0.15*sin(\x*360*1/0.75)-0.35});
        \draw[wave] plot[domain=-0.75:0.75,samples=50]
            (\x, {0.15*sin(\x*360*1.5/0.75)-0.7});
        \draw[wave] plot[domain=-0.75:0.75,samples=50]
            (\x, {0.15*sin(\x*360*2/0.75)-1.05});
    \end{scope}
    \begin{scope}[shift={([xshift=-0mm,yshift=-8mm]R.center)}]
         \draw[wave] plot[domain=-0.75:0.75,samples=50]
            (\x, {0.15*sin(\x*360*0.5/0.75)});
        \draw[wave] plot[domain=-0.75:0.75,samples=50]
            (\x, {0.15*sin(\x*360*1/0.75)-0.35});
    \end{scope}

    \begin{scope}[on background layer]
        \node[fill=mygreen!15, rounded corners,
              fit=(F)(R)(Finv), inner sep=10pt] (band) {};
    \end{scope}

    \node[blue] (W) [right=8mm of R] {W};

    \node[red] (plus)  [below=of Finv] {$+$};
    \node[red] (sigma) [below=of plus]      {$\sigma$};

    \draw[myarrow] (v)    -- (F);
    \draw[myarrow] (F)    -- (R);
    \draw[myarrow] (R)    -- (Finv);

    \draw[myarrow] (v.south) to[out=-30, in=90] (W.north);

    \draw[myarrow] (Finv) -- (plus);
    \draw[myarrow] (W) to[out=-90, in=0] (plus.east);

    \draw[myarrow] (plus) -- (sigma);

    \node[font=\large, anchor=west] at ([xshift=-1mm,yshift=2mm]v.north east) {Fourier layer};

    \begin{scope}[on background layer]
        \node[draw=black, rectangle, rounded corners, inner sep=12pt,
              fit=(v)(band)(W)(plus)(sigma)] (detailbox) {};
    \end{scope}

\end{scope}

\draw[dashed] (f2.north east) -- (detailbox.north west); %
\draw[dashed] (f2.south east) -- (detailbox.south west); %

\end{tikzpicture}

\caption{Architecture of the Fourier Neural Operator (FNO), adapted from \citep{li2020fno}. The left pipeline maps the input function $a(x)$ through a lifting operator $P$, a sequence of $T$ Fourier layers, and a projection $Q$ to the output $u(x)$. Each Fourier layer (detail, right) combines a spectral path  ($\mathcal{F} \to R \to \mathcal{F}^{-1}$) with a linear transformation $W$, followed by a nonlinearity $\sigma$.
}
\label{fig:FNO}

\end{figure}
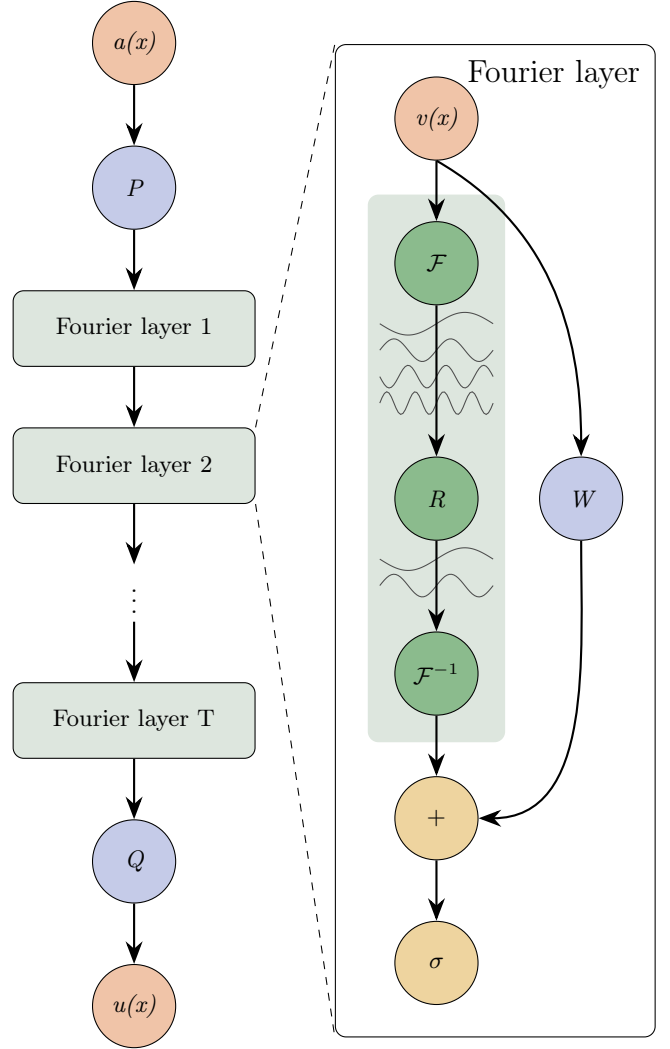

Neural operators compose linear integral operators $\mathcal{K}$ with pointwise non-linear activation functions~$\sigma$ to approximate highly non-linear operators. More precisely, we define the neural operator as
\begin{equation}
\label{eq:neural_operator}
\mathcal{G}_{\theta}
:=
\mathcal{Q}
\circ \mathcal{L}_{L}
\circ \cdots
\circ \mathcal{L}_{1}
\circ \mathcal{P},
\end{equation}
where each layer is given by
\begin{equation}
\label{eq:no_layer}
\mathcal{L}_{l}
=
\sigma\!\left(W_l+\mathcal{K}_l+b_l\right).
\end{equation}
Here, \(\mathcal{P}\) and \(\mathcal{Q}\) are pointwise neural networks that encode the lower-dimensional function into a higher-dimensional space and vice versa. The model stacks $L$ layers of $\mathcal{L}_l=\sigma(W_l+\mathcal{K}_l+b_l)$, where \(W_l\) are pointwise linear operators, \(\mathcal{K}_l\) are integral kernel operators, \(b_l\) are bias terms, and \(\sigma\) is a fixed activation function. The parameters $\theta$ consist of all parameters in $\mathcal{P}$, $\mathcal{Q}$, $W_l$, $\mathcal{K}_l$, and $b_l$. \citet{kossaifi2024neural} maintain a comprehensive open-source library for learning neural operators in PyTorch, which serves as the foundation for our implementation.

\label{appx:FNO-architecture}
\subsection{The FNO Architecture}

Throughout our numerical experiments, we employ Fourier neural operators to approximate the solution operators of PDEs. A \textbf{Fourier neural operator (FNO)}~\citep{li2020fno} is a neural operator using Fourier integral operator layers, which are defined via
\begin{equation}
\label{eq:Fourier}
\bigl(\mathcal{K}(\phi)v_t\bigr)(x)
=
\mathcal{F}^{-1}
\Bigl(
R_\phi \cdot (\mathcal{F} v_t)
\Bigr)(x).
\end{equation}
where $R_\phi$ is the Fourier transform of a periodic function $\kappa$ parameterized by \(\phi\). On a uniform mesh, the Fourier transform $\mathcal{F}$ can be implemented using the fast Fourier transform (FFT). Note that Fourier neural operators are differentiable. A depiction of the Fourier Neural Operator is shown in Fig.~\ref{fig:FNO}.

\label{appx:FNO-hps}
\subsection{ FNO Hyperparameters}
All Hamiltonian blocks mentioned in Appendix \ref{appx:block-diagonal} were trained using the same FNO architecture and hyperparameters. The details are summarized in
Table~\ref{tab:fno_training_hyperparameters}. The checkpoint with the
lowest validation loss was retained for subsequent surrogate evaluation and
inverse-design calculations.

\begin{table}[t]
\centering
\caption{FNO training hyperparameters.}
\label{tab:fno_training_hyperparameters}
\begin{tabular}{ll}
\hline
Hyperparameter & Value \\
\hline

Fourier modes & 60 \\
Hidden channels & 256 \\
Fourier layers & 4 \\
Lifting/projection ratio & 5 \\
Tensor factorization & Tucker \\
Tensor rank & 0.8 \\
Domain padding & 0.1 \\
Batch size & 32 \\
Training epochs & 200 \\
Optimizer & AdamW \\
Learning rate & \(5 \times 10^{-4}\) \\
Weight decay & \(3 \times 10^{-4}\) \\
Learning-rate scheduler & StepLR \\
Scheduler step size & 5 epochs \\
Scheduler decay factor & 0.7 \\
Training initial states & 1000 \\
Validation initial states & 100 \\
Detuning cutoff & \(10^{4}\) \\
Embedding scale & \(s_{\mathrm{emb}} = 0.05\) \\
Training seed & 1234 \\
Validation seed & 1 \\
\hline
\end{tabular}
\end{table}

\section{Additional Material}

\label{appx:RL}
\subsection{Reinforcement-learning benchmark details}

The RL benchmark was adapted from the RL-qMDP method in
Ref.~\citep{RLpipi}. The action library, referred to as the RL grid in
Fig.~\ref{fig:inverse_results}, contained $35280$ actions. This corresponds to
$1764$ frequency choices per polarization, two polarizations
$\sigma^+$ and $\sigma^-$, and $10$ pulse-duration indices,
\[
    \tau \in \{150,155,\ldots,195\}.
\]
The frequencies were restricted to the numerical frequency grid used for the
FNO-based comparison,
\[
   \omega/2\pi \in [5050,5300]~\mathrm{kHz}
\]
in the internal angular-frequency units of the numerical simulator. The
additional numerical pulses from the THz region were also included in the action
library. Thus, the RL agent and the FNO stochastic planner were evaluated on the
same discrete control grid for the benchmark comparison.

The Q-network had hidden dimension $128$ and was trained using replay batches
of size $256$, replay memory size $2.5\times 10^5$, discount factor
$\gamma=1$, Huber loss, and qMDP probability-weighted Bellman updates.
Exploration used an $\epsilon$-greedy policy with $\epsilon$ decayed from
$1.0$ to the specified final value $\epsilon_{\mathrm{end}}$ over
$7.2\times 10^4$ environment steps.
\begin{table}[h]
\centering
\caption{Sampled qMDP-DQN hyperparameter configurations used for the random-initialized RL benchmark. The dagger marks the selected policy used for the Monte-Carlo evaluation.}
\label{tab:rl_hyperparameter_sweep}
\begin{tabular}{ccccc}
\hline
Configuration &
$\tau_{\mathrm{RL}}$ &
$\eta_{\mathrm{RL}}$ &
$\rho$ &
$\epsilon_{\mathrm{end}}$ \\
\hline
0  & $1.0\times10^{-4}$ & $5.0\times10^{-4}$ & $1$ & $0.0125$ \\
1  & $2.5\times10^{-4}$ & $5.0\times10^{-4}$ & $2$ & $0.025$ \\
2  & $5.0\times10^{-4}$ & $5.0\times10^{-4}$ & $5$ & $0.005$ \\
\textbf{3}$^\dagger$ &
$\mathbf{1.0\times10^{-4}}$ &
$\mathbf{5.0\times10^{-4}}$ &
$\mathbf{2}$ &
$\mathbf{0.025}$ \\
4  & $5.0\times10^{-4}$ & $5.0\times10^{-4}$ & $1$ & $0.0125$ \\
5  & $2.5\times10^{-4}$ & $2.5\times10^{-4}$ & $2$ & $0.005$ \\
6  & $5.0\times10^{-4}$ & $5.0\times10^{-4}$ & $2$ & $0.0125$ \\
7  & $2.5\times10^{-4}$ & $1.0\times10^{-4}$ & $5$ & $0.005$ \\
8  & $2.5\times10^{-4}$ & $5.0\times10^{-4}$ & $1$ & $0.025$ \\
9  & $2.5\times10^{-4}$ & $1.0\times10^{-3}$ & $1$ & $0.025$ \\
10 & $5.0\times10^{-4}$ & $1.0\times10^{-3}$ & $2$ & $0.0125$ \\
11 & $1.0\times10^{-4}$ & $2.5\times10^{-4}$ & $2$ & $0.005$ \\
\hline
\end{tabular}
\end{table}
We performed a sampled hyperparameter sweep over the target-network update
rate $\tau_{\mathrm{RL}}$, learning rate $\eta_{\mathrm{RL}}$,
repeat-penalty weight $\rho$, and final exploration rate
$\epsilon_{\mathrm{end}}$. The full grid was
\[
\begin{aligned}
    \tau_{\mathrm{RL}} &\in
    \{1.0\times10^{-4},\,2.5\times10^{-4},\,5.0\times10^{-4}\}, \\
    \eta_{\mathrm{RL}} &\in
    \{1.0\times10^{-4},\,2.5\times10^{-4},\,5.0\times10^{-4},\,1.0\times10^{-3}\}, \\
    \rho &\in
    \{1,\,2,\,5\}, \\
    \epsilon_{\mathrm{end}} &\in
    \{0.005,\,0.0125,\,0.025\}.
\end{aligned}
\]
This gives $108$ possible configurations, from which $12$ configurations were
sampled using seed $100$. The sampled configurations are listed in
Table~\ref{tab:rl_hyperparameter_sweep}. Each configuration was trained for
$3000$ episodes. Policy snapshots were saved every $100$ episodes and evaluated
using $200$ Monte-Carlo rollouts. The selected policy was the snapshot at
episode $2900$ from configuration $3$. It was then evaluated on $1000$ Monte-Carlo trajectories using Monte-Carlo seed $777$, giving a
convergence rate of $0.428 \pm 0.016$ and an average episode length of
$48.974$ pulses.

\end{document}